\documentclass[final,1p,times]{elsarticle}

\journal{Journal of Computational Physics}

\usepackage{upgreek}

\usepackage{graphicx}
\usepackage{amsmath,amsfonts,amssymb,bm,times,mathtools}
\usepackage{pmbb-sym}
\usepackage{listings}

\usepackage{algorithm}
\floatname{algorithm}{Program}
\usepackage[noEnd]{algpseudocodex}
\algrenewcommand\algorithmicdo{}
\algrenewcommand\algorithmicthen{}
\algrenewcommand\algorithmicindent{0.8em}%

\usepackage{amsthm}

\newtheorem{remark}{Remark}

\newtheorem{assumption}{Assumption}

\usepackage{subfigure}
\usepackage[colorlinks,bookmarksopen,bookmarksnumbered,citecolor=red,urlcolor=red,breaklinks=true]{hyperref}

\usepackage{tikz}
\usetikzlibrary{positioning,shapes,arrows,arrows.meta}

\usepackage{here}

\usepackage{pgfplots}

\usepackage{amsmath,amsfonts,amssymb,bm,times,mathtools}

\def\abs#1{\left|{#1}\right|}
\def\bm#1{\boldsymbol{#1}}

\def\Order{\mathcal{O}}

\def\imath{\textrm{i}}

\usepackage{siunitx}
\NewDocumentCommand\numprint{m}{\num[round-mode = places]{#1}}
\NewDocumentCommand\nprounddigits{m}{\sisetup{round-precision = #1}}
\def\npproductsign#1{}
\def\Ns{N_{\mathrm{s}}}
\def\Nt{N_{\mathrm{t}}}
\def\Ne{N_{\mathrm{e}}}

\def\Ds{\Delta_{\mathrm{s}}}
\def\Dt{\Delta_{\mathrm{t}}}
\def\Hs{h_{\mathrm{s}}}
\def\Ht{h_{\mathrm{t}}}

\def\ds{d_{\mathrm{s}}}
\def\dt{d_{\mathrm{t}}}
\def\Mt{M_{\mathrm{t}}}
\def\Lag{\ell}

\def\Ps{{p_{\mathrm{s}}}}
\def\Pt{{p_{\mathrm{t}}}}

\def\maxlev{{l_{{\mathrm{max}}}}}
\def\Nleaf{N_{{\mathrm{leaf}}}}

\def\IL{{\mathcal I}}
\def\NL{{\mathcal N}}

\def\diff{\mathrm{d}}

\def\Order{\mathcal{O}}

\def\mat#1{\mathbf{#1}}

\def\rmO{O}
\def\rmS{S}
\def\rmI{I}
\def\rmJ{J}

\def\cO{\overline{\rmO}}
\def\cS{\overline{\rmS}}
\def\cI{\overline{I}}
\def\cJ{\overline{J}}

\def\Kop{{\mathcal K}}
\def\Lop{{\mathcal L}}
\def\Einc{\bm{E}^{\mathrm{I}}}
\def\Hinc{\bm{H}^{\mathrm{I}}}
\def\kinc{\bm{k}^{\mathrm{I}}}

\def\Ell{l}

\def\MFIE{\mathrm{MFIE}}
\def\DtEFIE{\partial_t\mathrm{EFIE}}
\def\DtMFIE{\partial_t\mathrm{MFIE}}

\begin{document}

\begin{frontmatter}
  
  \title{A fast time-domain boundary element method for three-dimensional electromagnetic scattering problems}
  
  \author[NU]{Toru Takahashi\corref{cor}}
  \ead{toru.takahashi@mae.nagoya-u.ac.jp}
  \cortext[cor]{Corresponding author}
  
  \address[NU]{Department of Mechanical Systems Engineering, Nagoya University, Furo-cho, Chikusa-ku, Nagoya city, Aichi, 464-8603 Japan}
  
  \begin{abstract}
    This paper proposes a fast time-domain boundary element method (TDBEM) to solve three-dimensional transient electromagnetic scattering problems regarding perfectly electric conductors in the classical marching-on-in-time manner. The algorithm of the fast TDBEM is a time-domain variant of the interpolation-based fast multipole method (IFMM), which is similar to the time-domain IFMM for acoustic scattering problems investigated in the author's previous studies. The principle of the present IFMM is to interpolate the kernel functions of the electric and magnetic field integral equations (EFIE and MFIE, respectively) so that every kernel function is expressed in a form of separation of variables in terms of both the spatial and temporal variables. Such an expression enables to construct a fast method to evaluate the scalar and vector potentials in the EFIE and MFIE with using so-called multipole-moments and local-coefficients associated with a space-time hierarchy. As opposed to $\Order(\Ns^2\Nt)$ of the conventional TDBEM, the computational complexity of the fast TDBEM is estimated as $\Order(\Ns^{1+\delta}\Nt)$, where $\Ns$ and $\Nt$ stand for the spatial and temporal degrees of freedom, respectively, and $\delta$ is typically $1/2$ or $1/3$. The numerical examples presented the advantages of the proposed fast TDBEM over the conventional TDBEM when solving large-scale problems.
  \end{abstract}
  
  
\end{frontmatter}

\section{Introduction}\label{s:intro}

Full-wave electromagnetic (EM) simulators are strong driving forces to produce scientific and industrial achievements nowadays, e.g. \cite{antonini2022} and the references therein. To deal with not only transient response but also frequency response (with the help of Fourier analysis), time-domain methods are useful. As such a method, the most widespread tool is presumably the finite-difference time-domain (FDTD) method~\cite{taflove1995computational}. On the other hand, the time-domain integral equation (IE) methods or, equivalently, the time-domain boundary element methods (TDBEMs) are rigorous and historical methods~\cite{stanley1979methods,miller1987overview,van2007electromagnetic,hargreaves2007phd,langer2008}, but their high computational cost as well as high memory consumption is often a severe problem in practical situations~\cite{chew2001book}.

To overcome the problem, the plane-wave time-domain (PWTD) algorithm was proposed as the fast algorithm for marching-on-in-time (MOT) based IE solvers firstly for acoustics by Ergin et al~\cite{ergin1998,ergin1999c,ergin2000} and then extended to electromagnetics by Shanker et al~\cite{shanker2000,shanker2003}. The PWTD can be esteemed as a time-domain version of the fast multipole method (FMM)~\cite{greengard1987,darve2000,nishimura2002,liu2009fast}. This is because the PWTD relies on (i) a far-field approximation with the help of the plane-wave expansion, which is utilised to represent the integral kernel (i.e. the fundamental solution for the wave equation) in a form of the separation of variables, and (ii) a hierarchical structure of computing domain, which is used to define far- and neighbour-fields in space-time. While the computational complexity of the conventional algorithm of the TDBEM is $\Order\left(\Ns^2\Nt\right)$ (where $\Ns$ and $\Nt$ stand for the spacial and temporal degrees of freedom, respectively), the PWTD realises a lower complexity of $\Order\left(\Ns^{1.5}(\log\Ns)\Nt\right)$ for the two-level approach~\cite{shanker2000} and further $\Order\left(\Ns(\log^2\Ns)\Nt\right)$ for the multi-level approach~\cite{shanker2003}. It is worth mentioning that the PWTD-accelerated time-domain IE solvers were parallelised to solve very large-scale problems of $\Ns\sim 10^6$ and $\Nt\sim 10^3$ on a memory-distributed computing system recently~\cite{liu2020parallel_PWTD}.

Following the studies on the PWTD~\cite{ergin1998,ergin1999c,ergin2000,shanker2000,shanker2003,aygun2004}, the present author proposed another fast algorithm for the TDBEM for 3D acoustics~\cite{takahashi2014}. The algorithm is similar to the PWTD, but the starting point is not the aforementioned plane-wave expansion but interpolation with respect to both spatial and temporal variables of the fundamental solution. The interpolated fundamental solution is in a form of separation of variables and, thus, allows to construct an FMM-like algorithm. This kind of FMM is termed the interpolation-based FMM (IFMM) whether it is for time domain~\cite{tausch2007,messner2015} or not~\cite{schobert2012}. Further, the recent work~\cite{takahashi2022} enhanced the formulation of \cite{takahashi2014} so that it can handle the B-spline function of not only the first order but also any order $d$ ($\ge 0$) as the temporal basis. The considerable difference from \cite{takahashi2014} is in the scheme of multipole-to-local (M2L) translation. The complexity of the IFMM-accelerated TDBEM for acoustics~\cite{takahashi2014,takahashi2022} is calculated as $\Order\left(\Ns^{1+\delta}\Nt\right)$, where $\delta$ is $1/3$ when boundary elements (to discretise the surface of sound scatterers) are arranged uniformly in a computational domain and $1/2$ when they are lying on a plane. As explained below, this computational complexity of the acoustic case is directly succeeded to the present EM case. Accordingly, the IFMM that the present study develops is semi-fast in comparison with the (multi-level) PWTD~\cite{shanker2003}. However, it should be emphasised that the numerical implementation of the IFMM is relatively facile because of its mathematical conciseness.

Only the PWTD~\cite{shanker2000,shanker2003} and the IFMM are not the accelerating methods to solve the IEs for electromagnetics. The other methods include the hierarchical FFT algorithm (HIL-FFT)~\cite{yilmaz2002HILFFT}, the time-domain adaptive integral method (TD-AIM)~\cite{yilmaz2004time_domain_AIM}, the two-level nonuniform grid time domain (NGTD) algorithm~\cite{boag2006NGTD} and its multi-level algorithm~\cite{liu2020parallel_PWTD}, the time-domain UV method~\cite{wang2011UV}, the fast dipole method~\cite{ding2013fast_dipole_method}, and potentially the accelerated Cartesian expansions (ACE) for evaluating general pairwise interactions~\cite{vikram2007ACE,shanker2007ACE}. The methodological and highest similarity of the proposed IFMM to ACE~\cite{vikram2007ACE} is to utilise the Taylor expansion fundamentally when both methods formulate their FMMs; precisely speaking, the proposed method exploits an interpolation of the kernel function to realise its separation of variables and then the Taylor expansion to construct a fast M2L operation, which is described in Algorithm 4 in \cite{takahashi2022}. However, extending the basic FMM operations presented in \cite[Theorems 2.3--2.5]{vikram2007ACE} from the case of a single target time-step to that of multiple target time-steps is not obvious. Hence, although the proposed method has a higher computational complexity of $\Order(\Ns^{1+\delta}\Nt)$ (where $\delta=1/3$ or $1/2$) than $\Order(\Ns\Nt)$ of ACE, the proposed IFMM is more general in the sense that it can deal with multiple target time-steps by way of the space-time hierarchy (see Eq.~(\ref{eq:eval})). Further, the present paper highlights a construction of the IFMM for electromagnetic integral equations rather than the particle simulations that ACE handles.

The purpose of this study is to progressively extend the IFMM for acoustic scattering problems~\cite{takahashi2014,takahashi2022} to EM scattering problems regarding perfectly electric conductors (PECs) in 3D. The acoustic case handles scalar fields, while the EM case does vector fields. This difference makes it attractive and challenging to construct the IFMM for the EM case. However, the overall algorithm remains almost the same as the acoustic case. Therefore, the prime focus of this article is on the formulation of the series of the basic operations necessary for building an FMM-like algorithm, i.e. (I) creation of multipole-moments (MMs), (II) translation of MMs to local-coefficients (LCs), and (III) evaluation of potentials by using LCs. In what follows, these operations are called P2M, M2L, and L2P, respectively, in accordance with the usual terminology of the FMM.

Selecting an IE to be solved is important. To eliminate the nonphysical resonance solution, it has been usual to use the combined field integral equation (CFIE) since Shanker et al~\cite{shanker2000cfie} established the CFIE, which was defined as a linear combination of the electrical field integral equation (EFIE) and the magnetic field integral equation (MFIE), upon the earlier works by Rynne et al~\cite{rynne1990stability} and Vechinski et al~\cite{vechinski1992cfie}. However, as indicated by Shanker et al~\cite{shanker2000cfie}, even the CFIE can be inaccurate owing to the employing numerical schemes (such as the \textit{explicit} MOT scheme used in \cite{vechinski1992cfie}) from the perspective of perturbations of the poles, which no longer correspond to those of the cavity resonance modes but those of the exterior problem in concern, of integral operators in Laplace-transformed domain~\cite{stanley1979methods}. Consequently, in the present case of the exterior scattering problems regarding PECs with using the \textit{implicit} MOT scheme, three types of the CFIE have been used under a variety of numerical schemes in the literature. The first type is the original one~\cite{shanker2000cfie}, i.e. a linear combination of the EFIE and the MFIE, and used in some works~\cite{shanker2000,wang2011UV,wang2021stable}. The second type is a combination of the temporal-derivatives of EFIE and that of MFIE, which are denoted by $\DtEFIE$ and $\DtMFIE$, respectively, hereafter. In particular, this type was selected to construct the multi-level PWTD~\cite{shanker2003}. The third type relies on the $\DtEFIE$ and the MFIE~\cite{jung2003}.

Dissimilarly to the above three types of CFIE, the present study employs another CFIE that comprises of $\DtEFIE$, $\DtMFIE$, and MFIE. This unconventional CFIE can indeed work properly under the presented numerical settings, i.e. the  Rao--Wilton--Glisson (RWG)~\cite{rao1982,rao1991transient} basis for space and the quadratic B-spline basis for time, as seen in the numerical test in Section~\ref{s:num}. In particular, the numerical result indicates that combining the three IEs is inevitable when the IFMM is used, in other words, when additional approximations are introduced.

It should be noted that using $\DtEFIE$ instead of the EFIE is necessary from the viewpoint of the computational cost. The naive EFIE contains the temporal integral in its scalar potential. Hence, unless the temporal integral is erased by the temporal differentiation, the cost to solve the EFIE (even combined with MFIE and/or $\DtMFIE$) scales as $O(\Ns^2\Nt^2)$ in the conventional algorithm. It is possible to construct an IFMM for the EFIE, but the resulting TDBEM would be slower than an IFMM based on $\DtEFIE$. This is because the cancellation of the successive discretised potentials (see (\ref{eq:vanish})) does not hold for the scalar potential in the EFIE. However, using the Hertz vector $\bm{P}$ such that $\partial_t\bm{P}=\bm{J}$ can avoid the temporal differentiation of the EFIE~\cite{jung2003,chung2003solution} and enables to construct an IFMM in the same way as the case of $\DtEFIE$.

The remainder of this article is organised as follows: Section~\ref{s:tdbem} shows the formulation of the TDBEM based on the aforementioned CFIE and the discretisation scheme. Section~\ref{s:fmm} constructs the IFMM to speed up the TDBEM in the previous section. The emphasis is on the formulation of the aforementioned operations (I)--(III) by deriving the expression of the scalar/vector potentials with the MMs and LCs in a general space-time configuration. Section~\ref{s:num} assesses the proposed fast TDBEM numerically with regard to the accuracy, performance, memory usage, and applicability. Section~\ref{s:conclusion} concludes the present study.

\section{TDBEM for EM scattering problems}\label{s:tdbem}

This section describes the conventional TDBEM based on the aforementioned CFIE, which consists of $\DtEFIE$, $\DtMFIE$, and MFIE, regarding PECs irradiated by an incident EM wave. The CFIE is discretised with the RWG basis~\cite{rao1982} for space and the B-spline basis for time. As usual, the Galerkin method is used for space, which helps to reduce the singularity of the scalar potential of the $\DtEFIE$. Last, the discretised CFIE is solved in a standard MOT scheme.

\subsection{Problem statement and CFIE}\label{s:tdbem_problem}

\begin{figure}[hbt]
  \centering
  \includegraphics[width=.4\textwidth]{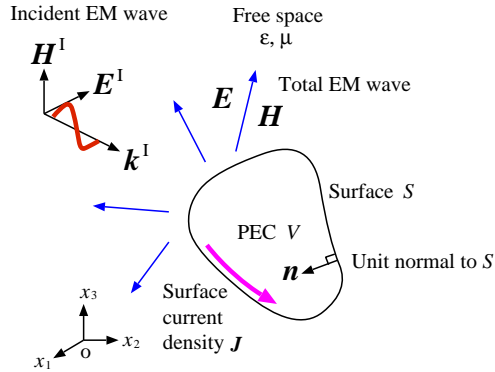}
  \caption{EM scattering problems. The vector $\kinc$ denotes the propagating direction of the incident EM wave.}
  \label{fig:problem}
\end{figure}

As illustrated in Figure~\ref{fig:problem}, let PECs exist in a domain $V$ in the infinite (free) space $\bbbr^3$, where the permittivity and permeability are given by $\epsilon$ and $\mu$, respectively; then, the wave speed $c$ and the wave impedance $\eta$ are determined as $(\epsilon\mu)^{-1/2}$ and $(\mu/\epsilon)^{1/2}$, respectively. The surface or boundary $\partial V$ is denoted by $S$. The unit normal vector $\bm{n}$ to $S$ is supposed to be directed to the inside of $V$, following the usual manner in the field of the BEM for exterior problems. When incident EM fields, denoted by $\Einc$ and $\Hinc$, are given, the problem is to solve the induced surface current density $\bm{J}:=\bm{n}\times\bm{H}$, which is supposed to be zero when the time $t\le 0$, from the following CFIE:
\begin{eqnarray}
  &&\dot{\Lop}(\bm{J})(\bm{x},t)+\theta\eta\left(\dot{\Kop}(\bm{J})(\bm{x},t)+\zeta\Kop(\bm{J})(\bm{x},t)\right)\nonumber\\
  &&=\bm{n}(\bm{x})\times\bm{n}(\bm{x})\times\dot{\Einc}(\bm{x},t)+\theta\eta\bm{n}(\bm{x})\times\left(\dot{\Hinc}(\bm{x},t)+\zeta\Hinc(\bm{x},t)\right)\quad\bm{x}\in S,\ t>0,
  \label{eq:cfie}
\end{eqnarray}
where $\theta$ ($>0$) and $\zeta$ ($>0$) denote predefined coupling parameters, the dot $\dot{(\ )}$ denotes the temporal differentiation $\partial_t$, and the operators $\Lop$ and $\Kop$ correspond to those of the standard EFIE and MFIE, respectively, i.e.
\begin{eqnarray}
  \dot{\Lop}(\bm{J})(\bm{x},t)&:=&-\bm{n}(\bm{x})\times\bm{n}(\bm{x})\times\left(\ddot{\bm{A}}(\bm{x},t)+\nabla_x\dot{\phi}(\bm{x},t)\right),\\
  \Kop(\bm{J})(\bm{x},t)&:=&\bm{n}(\bm{x})\times\bm{P}(\bm{x},t).
\end{eqnarray}
Here, the scalar potential $\phi$ and the vector potentials $\bm{A}$ and $\bm{P}$ are defined as follows:
\begin{subequations}
  \begin{eqnarray*}
    \bm{A}(\bm{x},t)&:=&\frac{\mu}{4\pi}\int_S\frac{\bm{J}(\bm{y},t-r/c)}{r}\diff S_y,\nonumber\\
    \phi(\bm{x},t)&:=&-\frac{1}{4\pi\varepsilon}\int_S \int_0^{(t-r/c)_+} \frac{\nabla_y\cdot\bm{J}(\bm{y},s)}{r} \diff s \diff S_y,\nonumber\\
    \bm{P}(\bm{x},t)&:=&\frac{1}{4\pi}\int_S\nabla_x\times\frac{\bm{J}(\bm{y},t-r/c)}{r}\diff S_y
  \end{eqnarray*}%
\end{subequations}
and consequently
\begin{subequations}
  \begin{eqnarray}
    &&\ddot{\bm{A}}(\bm{x},t)=\frac{\mu}{4\pi}\int_S\frac{\ddot{\bm{J}}(\bm{y},t-r/c)}{r}\diff S_y,\label{eq:def_A}\\
    &&\dot{\phi}(\bm{x},t)=-\frac{1}{4\pi\varepsilon}\int_S \frac{\nabla_y\cdot\bm{J}(\bm{y},t-r/c)}{r} \diff S_y, \label{eq:def_phi}\\
    &&\dot{\bm{P}}(\bm{x},t)+\zeta\bm{P}(\bm{x},t)=\frac{1}{4\pi}\int_S\nabla_x\times\frac{\dot{\bm{J}}(\bm{y},t-r/c)+\zeta\bm{J}(\bm{y},t-r/c)}{r}\diff S_y. \label{eq:def_P}
  \end{eqnarray}%
  \label{eq:layer}%
\end{subequations}
In these, $r$ denotes the distance between two points $\bm{x}$ and $\bm{y}$, i.e. $r:=|\bm{x}-\bm{y}|$, and $(\cdot)_+$ denotes the truncated power function (of degree one), i.e. $(x)_+=0$ if $x<0$ and $(x)_+=x$ if $x\ge0$.

\subsection[Approximation of current density]{Approximation of $\bm{J}$}

\begin{figure}[hbt]
  \centering
  \includegraphics[width=.2\textwidth]{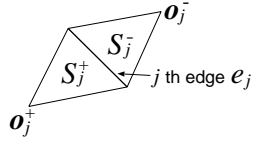}
  \caption{Notations for the $j$th RWG basis function $\bm{f}_j$.}
  \label{fig:rwg-basis}
\end{figure}

The surface $S$ is discretised with $\Ns$ planar triangular elements, denoted by $S_1,\ldots,S_{\Ns}$. In addition, equidistant time-steps are introduced, i.e. $t_\beta:=\beta\Dt$ ($\beta=0,1,2...$) denotes the $\beta$th time-step and $\Dt$ denotes the prescribed time-step length. Then, the surface current density $\bm{J}$ in (\ref{eq:cfie}) is approximated as follows:
\begin{eqnarray}
  \bm{J}(\bm{x},t) \approx \sum_{j=1}^{\Ne} \sum_{\beta\ge 0} J_j^\beta \bm{f}_j(\bm{x}) N^{\beta,d}(t) \quad \bm{x}\in S,\ t>0,
  \label{eq:J}
\end{eqnarray}
where $\Ne$ stands for the number of edges, denoted by $e_1,\ldots,e_{\Ne}$, of all the triangles and $\bm{f}_j$ represents $e_j$'s RWG basis function~\cite{rao1982} defined by
\begin{eqnarray*}
  \bm{f}_j(\bm{x}) := \begin{cases}
    \frac{\pm\Ell_j}{2A_j^\pm}(\bm{x}-\bm{o}_j^\pm) & \mbox{if $\bm{x}\in S_j^\pm$} \\
    \bm{0} & \mbox{otherwise}
  \end{cases}.
\end{eqnarray*}
Here, $S_j^+$ (respectively, $S_j^-$), whose area is $A_j^+$ ($A_j^-$), stands for the triangles in the positive (negative) side of $e_j$, whose length is $\Ell_j$; see Figure~\ref{fig:rwg-basis}. In addition, $\bm{o}_j^\pm$ denotes $S_j^\pm$'s vertex that is not on $e_j$. Also, $N^{\beta,d}$ denotes the $d$-th order B-spline basis for time and can be split to $d+2$ truncated power functions of degree $d$ as follows~\cite{takahashi2022}:
\begin{eqnarray*}
  N^{\beta,d}(t) = \sum_{\kappa=0}^{d+1}w^{\kappa,d}\left(\frac{t-t_{\beta+\kappa}}{\Dt}\right)^d_+
  \quad\beta=0,1,2,\ldots.
\end{eqnarray*}
Here, the weight $w^{\kappa,d}$ is given as 
\begin{eqnarray*}
  w^{\kappa,d}=\frac{d+1}{\Pi_{i=0;i\ne\kappa}^{d+1}(i-\kappa)}\quad \kappa=0,\ldots,d+1.
\end{eqnarray*}

In the previous study on the construction of the IFMM for transient acoustic scattering problems~\cite{takahashi2022}, the Burton--Miller integral equation (BMIE) is targeted. The BMIE, which corresponds to the CFIE in (\ref{eq:cfie}), consists of the single- and double-layer potentials and their normal derivatives, which are scalar-valued functions. In addition, to approximate the boundary variables (i.e. sound pressure and its normal derivative) in the BMIE, the spatial basis is chosen as the piece-wise constant basis differently from the RWG basis, while the temporal basis is the aforementioned B-spline basis. From these, the discretisation of the scalar/vector potentials in (\ref{eq:layer}) (and also the formulation of the IFMM investigated in the next section) is relatively complicated, but the way of thinking is the same as the acoustic case.

Further, the BMIE employs the collocation method, while the CFIE does the Galerkin method. However, the role of the IFMM is to evaluate potentials at certain points $\bm{x}\in S$. Hence, the integral with respect to $\bm{x}\in S$, which appears after the testing procedure of the CFIE, does not matter when the IFMM is constructed for the EM case.

\subsection{Discretisation of the the potentials in the CFIE}\label{s:tdbem_disc}

Using the approximation in (\ref{eq:J}), the scalar and vector potentials in (\ref{eq:layer}) can be discretised at $t=t_\alpha$ (where $\alpha=1,\ldots,\Nt-1$) as follows:
\begin{subequations}
  \begin{eqnarray}
    &&\ddot{\bm{A}}(\bm{x},t_\alpha)
    = \sum_{j=1}^{\Ne} \sum_{\beta=0}^{\alpha-1} \sum_{\kappa=0}^{d+1} w^{\kappa,d} \underbrace{\frac{\mu}{4\pi}\sum_{\sigma=\pm}\frac{\sigma\Ell_j}{2A_j^\sigma (c\Dt)^d}\int_{S_j^\sigma} \ddot{U}^d(\bm{x},\bm{y},t_\alpha,t_{\beta+\kappa})(\bm{y}-\bm{o}_j^\sigma)\diff S_y}_{\displaystyle=:\ddot{\bm{A}}_j^{(\alpha-\beta-\kappa)}(\bm{x})} J_j^\beta,\\
    &&\dot{\phi}(\bm{x},t_\alpha)
    = \sum_{j=1}^{\Ne} \sum_{\beta=0}^{\alpha-1} \sum_{\kappa=0}^{d+1} w^{\kappa,d} \underbrace{\frac{-1}{4\pi\varepsilon} \sum_{\sigma=\pm} \frac{\sigma\Ell_j}{A_j^\sigma (c\Dt)^d} \int_{S_j^\sigma} U^d(\bm{x},\bm{y},t_\alpha,t_{\beta+\kappa}) \diff S_y}_{\displaystyle=:\dot{\phi}_j^{(\alpha-\beta-\kappa)}(\bm{x})} J_j^\beta,\\
    &&\dot{\bm{P}}(\bm{x},t_\alpha)+\zeta\bm{P}(\bm{x},t_\alpha)\nonumber\\
    &&= \sum_{j=1}^{\Ne} \sum_{\beta=0}^{\alpha-1} \sum_{\kappa=0}^{d+1} w^{\kappa,d} \underbrace{\frac{1}{4\pi} \sum_{\sigma=\pm} \frac{\sigma\Ell_j}{2A_j^\sigma(c\Dt)^d} \int_{S_j^\sigma} (\bm{y}-\bm{o}_j^\sigma) \times\nabla_y \left(\dot{U}^d(\bm{x},\bm{y},t_\alpha,t_{\beta+\kappa})+\zeta U^d(\bm{x},\bm{y},t_\alpha,t_{\beta+\kappa})\right)\diff S_y}_{\displaystyle=:\dot{\bm{P}}_j^{(\alpha-\beta-\kappa)}(\bm{x})+\zeta\bm{P}_j^{(\alpha-\beta-\kappa)}(\bm{x})} J_j^\beta,
  \end{eqnarray}%
  \label{eq:layer_disc_tmp}%
\end{subequations}
where the summation over $\sigma$ considers the positive and negative sides of the edge $e_j$. In addition, the kernel function $U^d$ is defined as follows:
\begin{eqnarray}
  U^d(\bm{x},\bm{y},t,s):=\frac{(c(t-s)-\abs{\bm{x}-\bm{y}})_+^d}{\abs{\bm{x}-\bm{y}}},
  \label{eq:U}
\end{eqnarray}
which is exactly the same as the kernel function used for the acoustic case~\cite[Eq.~(10)]{takahashi2022}.

It should be noted that the coefficients of $J_j^\beta$ in the RHSs of (\ref{eq:layer_disc_tmp}) are determined by the difference of $\alpha$ and $\beta+\kappa$ because of the definition of $U^d$ in (\ref{eq:U}) and the assumption of the equidistant time-steps. To emphasise this, the difference is written inside brackets.

Similarly to the acoustic case~\cite[Remark~4]{takahashi2022}, because the support of $N^{\beta,d}$ is finite, i.e. $[t_\beta,t_{\beta+d+1}]$, it can be proven that
\begin{eqnarray}
  \sum_{\kappa=0}^{d+1} w^{\kappa,d} \chi_{ij}^{(\gamma-\kappa)}(\bm{x})\equiv 0 \quad\text{if $\gamma>\gamma^*$},
  \label{eq:vanish}
\end{eqnarray}
where
\begin{eqnarray}
  \gamma^*:=\frac{\max_{\bm{x},\bm{y}\in S}\abs{\bm{x}-\bm{y}}}{c\Dt}+d+1
  \label{eq:gamma*}
\end{eqnarray}
is a constant determined by the geometry of $S$, $\Dt$, and $d$. Also, $\chi_{ij}^{(\gamma)}$ is the representative of $\bm{A}_{ij}^{(\gamma)}$, $\dot{\bm{A}}_{ij}^{(\gamma)}$, $\ddot{\bm{A}}_{ij}^{(\gamma)}$, $\dot{\phi}_{ij}^{(\gamma)}$, $\bm{P}_{ij}^{(\gamma)}$, and $\dot{\bm{P}}_{ij}^{(\gamma)}$. It should be noted that $\phi_{ij}^{(\gamma)}$ cannot have the vanishing property in (\ref{eq:vanish}) because of its temporal integral; from this reason, the present study does not utilise the EFIE but $\DtEFIE$.

Using (\ref{eq:vanish}), one can replace $\sum_{\beta=0}^{\alpha-1}$ in (\ref{eq:layer_disc_tmp}) with $\sum_{\beta=\beta^*}^{\alpha-1}$, where $\beta^*:=\alpha-\gamma^*$. As a result, the following expressions are obtained:
\begin{subequations}
  \begin{eqnarray}
    \ddot{\bm{A}}(\bm{x},t_\alpha)
    &=& \sum_{j=1}^{\Ne} \sum_{\beta=\beta^*}^{\alpha-1} \sum_{\kappa=0}^{d+1} w^{\kappa,d} \ddot{\bm{A}}_j^{(\alpha-\beta-\kappa)}(\bm{x}) J_j^\beta, \label{eq:layer_disc_tmp2_A}\\
    \dot{\phi}(\bm{x},t_\alpha)
    &=& \sum_{j=1}^{\Ne} \sum_{\beta=\beta^*}^{\alpha-1} \sum_{\kappa=0}^{d+1} w^{\kappa,d} \dot{\phi}_j^{(\alpha-\beta-\kappa)}(\bm{x}) J_j^\beta,\\
    \dot{\bm{P}}(\bm{x},t_\alpha)+\zeta\bm{P}(\bm{x},t_\alpha)
    &=& \sum_{j=1}^{\Ne} \sum_{\beta=\beta^*}^{\alpha-1} \sum_{\kappa=0}^{d+1} w^{\kappa,d} \left(\dot{\bm{P}}_j^{(\alpha-\beta-\kappa)}(\bm{x})+\zeta\bm{P}_j^{(\alpha-\beta-\kappa)}(\bm{x})\right) J_j^\beta.
  \end{eqnarray}%
  \label{eq:layer_disc_tmp2}%
\end{subequations}
It should be remarked that every summation over $\beta$ considers at most $\gamma^*-1$ terms. 

To simplify the above expressions, a new variable $\hat{J}_j$ is defined as a weighted sum of $J_j^\beta,\ldots,J_j^{\beta-d-1}$, i.e.
\begin{eqnarray}
  \hat{J}_j^\beta:=\sum_{\kappa=0}^{\min(d+1,\ \beta-\beta^*)}w^{\kappa,d}J_j^{\beta-\kappa}\quad \beta^*\le\beta\le\alpha-1.
  \label{eq:hat_J}
\end{eqnarray}
Moreover, $\sum_{\beta=\beta^*}^{\alpha-1}$ is written to $\sum_{\beta=\beta^*+1}^{\alpha}$ by re-defining $\beta+1$ as $\beta$. Then, as shown in \ref{s:simplify}, the potentials in (\ref{eq:layer_disc_tmp2}) can be written as follows:
\begin{subequations}
  \begin{eqnarray}
    \ddot{\bm{A}}(\bm{x},t_\alpha)
    &=& \sum_{j=1}^{\Ne} \sum_{\beta=\beta^*+1}^{\alpha} \ddot{\bm{A}}_j^{(\alpha-\beta+1)}(\bm{x}) \hat{J}_j^{\beta-1}, \label{eq:layer_disc_A}\\
    \dot{\phi}(\bm{x},t_\alpha)
    &=& \sum_{j=1}^{\Ne} \sum_{\beta=\beta^*+1}^{\alpha} \dot{\phi}_j^{(\alpha-\beta+1)}(\bm{x}) \hat{J}_j^{\beta-1}, \label{eq:layer_disc_phi}\\
    \dot{\bm{P}}(\bm{x},t_\alpha)+\zeta\bm{P}(\bm{x},t_\alpha)
    &=& \sum_{j=1}^{\Ne} \sum_{\beta=\beta^*+1}^{\alpha} \left(\dot{\bm{P}}_j^{(\alpha-\beta+1)}(\bm{x})+\zeta\bm{P}_j^{(\alpha-\beta+1)}(\bm{x})\right) \hat{J}_j^{\beta-1}. \label{eq:layer_disc_P}%
  \end{eqnarray}%
  \label{eq:layer_disc}%
\end{subequations}%

It is worth mentioning that the integrals in (\ref{eq:layer_disc}) as well as their spatial and temporal derivatives can be evaluated analytically. To this end, the integrals are written as follows:
\begin{subequations}
  \begin{eqnarray*}
    \ddot{\bm{A}}_j^{(\gamma)}(\bm{x}) &=& \frac{\mu}{4\pi} \sum_{\sigma=\pm} \frac{\sigma\Ell_j}{2A_j^\sigma(c\Dt)^d}\left( (\bm{x}-\bm{o}_j^\sigma) \ddot{I}_d(\bm{x},t_{\gamma}) - \ddot{\bm{I}}_d(\bm{x},t_{\gamma}) \right), \\
    \dot{\phi}_j^{(\gamma)}(\bm{x}) &=& \frac{-1}{4\pi\varepsilon} \sum_{\sigma=\pm} \frac{\sigma\Ell_j}{A_j^\sigma (c\Dt)^d} I_d(\bm{x},t_{\gamma}), \\
    \bm{P}_j^{(\gamma)}(\bm{x}) &=& \frac{1}{4\pi} \sum_{\sigma=\pm}\frac{\sigma\Ell_j}{2A_j^\sigma(c\Dt)^d}(\bm{x}-\bm{o}_j^\sigma)\times\left(-\nabla_x I_d(\bm{x},t_\gamma)\right),
  \end{eqnarray*}%
\end{subequations}
where the identity $(\bm{y}-\bm{o}_j^\sigma)\times\bm{r}\equiv(\bm{x}-\bm{o}_j^\sigma)\times\bm{r}$ was used in the last expression and the following scalar- and vector-valued integrals are defined:
\begin{eqnarray*}
  I_d(\bm{x},t) := \int_{S_j^\sigma} \frac{(ct-r)_+^d}{r} \diff S_y,\quad
  \bm{I}_d(\bm{x},t) := \int_{S_j^\sigma} \frac{(ct-r)_+^d}{r}(\bm{x}-\bm{y})\ \diff S_y \equiv - \nabla_x \int_{S_j^\sigma} \frac{(ct-r)_+^{d+1}}{d+1} \diff S_y.
\end{eqnarray*}
The way of analytical evaluation of $I_d$ is the same as the acoustic case~\cite[Appendix B]{takahashi2022}, while $\bm{I}_d$ can be calculated by performing the integral similarly to $I_d$ and then differentiating the result with respect to the evaluation point $\bm{x}$.

However, the IFMM evaluates the integrals numerically. This is relatively inaccurate but acceptable because $\bm{x}$ is sufficiently far from $\bm{y}$ by construction and, thus, those integrals are non-singular.

\subsection{Testing procedure}

The RWG basis $\bm{f}_i$, where $i=1,\ldots,\Ne$, is applied to the CFIE in (\ref{eq:cfie}) as the test function. Then, the tested terms in the LHS of (\ref{eq:cfie}) can be written from (\ref{eq:layer_disc}) as follows:
\begin{subequations}
  \begin{eqnarray}
    &&\int_{S} \bm{f}_i(\bm{x})\cdot\left(-\bm{n}(\bm{x})\times\bm{n}(\bm{x})\times\ddot{\bm{A}}(\bm{x},t_\alpha)\right) \diff S_x 
    =\int_{S} \bm{f}_i(\bm{x})\cdot\ddot{\bm{A}}(\bm{x},t_\alpha)\diff S_x 
    = \sum_{j=1}^{\Ne} \sum_{\beta=\beta^*+1}^{\alpha}\ddot{A}_{ij}^{(\alpha-\beta+1)} \hat{J}_j^{\beta-1},\\
    &&\int_{S} \bm{f}_i(\bm{x})\cdot\left(-\bm{n}(\bm{x})\times\bm{n}(\bm{x})\times\nabla_{S_x}\dot{\phi}(\bm{x},t_\alpha)\right)\diff S_x
    = - \int_{S} \left(\nabla_{S_x}\cdot\bm{f}_i(\bm{x})\right) \dot{\phi}(\bm{x},t_\alpha)\diff S_x
    = - \sum_{j=1}^{\Ne} \sum_{\beta=\beta^*+1}^{\alpha} \dot{\phi}_{ij}^{(\alpha-\beta+1)} \hat{J}_j^{\beta-1},\nonumber\\&&\\
    &&\int_{S} \bm{f}_i(\bm{x}) \cdot \bm{n}(\bm{x})\times\left(\dot{\bm{P}}(\bm{x},t_\alpha)+\zeta\bm{P}(\bm{x},t_\alpha)\right) \diff S_x
    = \sum_{j=1}^{\Ne}\sum_{\beta=\beta^*+1}^{\alpha} \left(\dot{P}_{ij}^{(\alpha-\beta+1)}+\zeta P_{ij}^{(\alpha-\beta+1)}\right) \hat{J}_j^{\beta-1},
  \end{eqnarray}%
  \label{eq:testing}%
\end{subequations}
where
\begin{eqnarray*}
  \ddot{A}_{ij}^{(\gamma)}
  &:=&\sum_{\rho=\pm}\int_{S_i^\rho} \bm{f}_i(\bm{x}) \cdot \ddot{\bm{A}}_{j}^{(\gamma)}(\bm{x})\ \diff S_x,\\
  \dot{\phi}_{ij}^{(\gamma)}
  &:=&\sum_{\rho=\pm}\int_{S_i^\rho} \left(\nabla_{S_x}\cdot\bm{f}_i(\bm{x})\right) \dot{\phi}_j^{(\gamma)}(\bm{x})\ \diff S_x,\\
  \dot{P}_{ij}^{(\gamma)}+\zeta P_{ij}^{(\gamma)}
  &:=&\sum_{\rho=\pm}\int_{S_i^\rho} \bm{f}_i(\bm{x}) \cdot \bm{n}(\bm{x})\times \left(\dot{\bm{P}}_j^{(\gamma)}(\bm{x})+\zeta \bm{P}_j^{(\gamma)}(\bm{x})\right)\diff S_x.
\end{eqnarray*}
Here, the domain of the integral over $\bm{x}$ is limited to $S_i^\pm$ because the support of $\bm{f}_i$ is $S_i^+\cup S_i^-$.From (\ref{eq:testing}), the CFIE in (\ref{eq:cfie}) can be discretised as follows:
\begin{eqnarray}
  \sum_{\beta=\beta^*+1}^{\alpha} \mat{Z}^{(\alpha-\beta+1)} \hat{\mat{J}}^{\beta-1} = \mat{V}^{\alpha-1}\qquad \alpha=1,\ldots,\Nt-1,
  \label{eq:system}
\end{eqnarray}
where $\mat{Z}^{(\gamma)}$ is the $\Ne\times\Ne$-dimensional coefficient matrix regarding a time-step difference of $\gamma$ ($\ge 1$) and its $(i,j)$th element is defined by
\begin{eqnarray}
  Z_{ij}^{(\gamma)} := \ddot{A}_{ij}^{(\gamma)}-\dot{\phi}^{(\gamma)}+\theta\eta\left(\dot{P}_{ij}^{(\gamma)}+\zeta P_{ij}^{(\gamma)}\right)
  \label{eq:z_ij}
\end{eqnarray}
for $i,j=1,\ldots,\Ne$. 
Also, $\hat{\mat{J}}^{\alpha-1}$ is the $\Ne$-dimensional vector whose $j$th element is $\hat{J}_j^{\alpha-1}$. In addition, $\mat{V}^{\alpha-1}$ is also the $\Ne$-dimensional vector and its $i$th component $V_i^{\alpha-1}$ is given by
\begin{eqnarray}
  V_i^{\alpha-1} := \sum_{\rho=\pm}\int_{S_i^\rho}\bm{f}_i(\bm{x}) \cdot \left[-\dot{\bm{E}}^\mathrm{I}(\bm{x},t_\alpha)
  +\theta\eta\bm{n}(\bm{x})\times\left(\dot{\bm{H}}^\mathrm{I}(\bm{x},t_\alpha)+\zeta\bm{H}^\mathrm{I}(\bm{x},t_\alpha)\right)\right]\diff S_x.
  \label{eq:v_i}
\end{eqnarray}

To evaluate each spatial integral on a triangle $S_i^\rho$ with respect to $\bm{x}$ in (\ref{eq:z_ij}) and (\ref{eq:v_i}), the Gaussian quadrature formula is applied with a predefined number of quadrature points. In the numerical experiments in Section~\ref{s:num}, three points formula is used.

\subsection{Solving the discretised CFIE}\label{s:tdbem_solve}

To solve the unknown vector $\mat{J}^{\alpha-1}$ at the current $\alpha$th time-step according to the MOT scheme, (\ref{eq:system}) is rewritten as 
\begin{eqnarray}
  w^{0,d}\mat{Z}^{(1)}\mat{J}^{\alpha-1}=\mat{R}^{\alpha-1} \qquad \alpha=1,\ldots,\Nt-1,
  \label{eq:solve}
\end{eqnarray}
where the RHS vector is defined as
\begin{eqnarray}
  \mat{R}^{\alpha-1}:=\mat{V}^{\alpha-1}- \sum_{\beta=\beta^*+1}^\alpha \mat{Z}^{(\alpha-\beta+1)}\hat{\mat{J}}^{\beta-1}.
  \label{eq:R}
\end{eqnarray}
Here, $\hat{\mat{J}}^{\beta}$ in (\ref{eq:hat_J}) should be redefined so that it excludes the underlying unknown vector $\mat{J}^{\alpha-1}$, i.e.
\begin{eqnarray*}
  \hat{\mat{J}}^{\beta} :=
  \begin{cases}
    \displaystyle\sum_{\kappa=0}^{\min(d+1,\ \beta-\beta^*)}w^{\kappa,d}\mat{J}^{\beta-\kappa} & \mbox{if $\beta^*\le\beta\le\alpha-2$} \\
    \displaystyle\sum_{\kappa=1}^{\min(d+1,\ \beta-\beta^*)}w^{\kappa,d}\mat{J}^{\beta-\kappa} & \mbox{if $\beta=\alpha-1$}
  \end{cases}.
\end{eqnarray*}

\subsection{Computational complexity}\label{s:tdbem_complexity}

The estimation of the computational complexity of the above conventional TDBEM is made on the following assumptions:
\begin{assumption}\label{assume:S}
  The size of the boundary $S$ is $\Order(1)$, i.e. the size is independent of both $\Ns$ ($\sim\Ne$) and $\Nt$.
\end{assumption}
\begin{assumption}\label{assume:Dt}
  The time-step size $\Dt$ is $\Order(1)$.
\end{assumption}
From these, $\gamma^*$ in (\ref{eq:gamma*}) is $\Order(1)$. Then, the computational cost to evaluate the RHS vector $\mat{R}^{\alpha-1}$ in (\ref{eq:R}), which is comprised of matrix-vector products $\mat{Z}^{(\alpha-\beta+1)}\hat{\mat{J}}^{\beta-1}$, is $\Order(\gamma^*\Ne^2)=\Order(\Ns^2)$ for each $\alpha$ because the matrix $\mat{Z}^{(\gamma)}$ becomes denser as $\gamma$ becomes larger. On the other hand, $\mat{Z}^{(1)}$ in the LHS of (\ref{eq:solve}) is close to diagonal and, thus, the cost to solve (\ref{eq:solve}) with respect to $\mat{J}^{\alpha-1}$ can be assumed to be $\Order(\Ne)=\Order(\Ns)$; the GMRES~\cite{saad1986GMRES} is utilised as the solver with using $10^{-8}$ as the relative drop tolerance in the numerical analyses in Section~\ref{s:num}. Therefore, the computational complexity over all the time steps is estimated as $\Order(\Ns^2\Nt)$, which can prohibit the conventional TDBEM from solving large-scale problems. 

The IFMM plays a role to approximately evaluate the summation of the matrix-vector products (except for the spatially-near interactions) in the RHS vectors in (\ref{eq:R}) with a computational cost less than $\Order(\Ns^2\Nt)$.

\section{Interpolation-based FMM}\label{s:fmm}

This section presents the fast TDBEM to solve the discretised CFIE in (\ref{eq:solve}). Sections~\ref{s:fmm_config}--\ref{s:fmm_cfie} formulate the fundamental three steps (I)--(III) mentioned in Section~\ref{s:intro}, i.e. P2M, M2L, and L2P, for the scalar and vector potentials of the CFIE in (\ref{eq:cfie}). From these as well as the M2M and L2L formulae, the algorithm of the (multi-level) fast TDBEM is constructed in Section~\ref{s:fmm_algo}. Successively, the computational complexity of the proposed fast method is estimated in Section~\ref{s:fmm_complexity}. Last, some notes on the numerical implementation are given in Section~\ref{s:fmm_note}.

\subsection{Configuration}\label{s:fmm_config}

\def\OI{(\rmO,\rmI)}
\def\SJ{(\rmS,\rmJ)}

\begin{figure}[H]
  \centering
  \includegraphics[width=.6\textwidth]{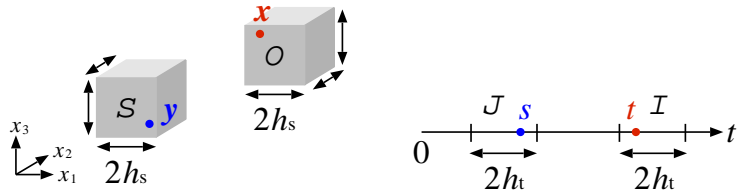}
  \caption{General configuration of two cells $\rmO$ and $\rmS$ in 3D (left) and two time-intervals $\rmI$ and $\rmJ$ on the time-axis (right) in order to formulate the fundamental operations of the IFMM.}
  \label{fig:config}
\end{figure}

The above three steps are formulated in a general configuration. Let $\rmO$ and $\rmS$ be cubic boxes (called \textit{cells} hereafter) with the edge length of $2\Hs$ in 3D (Figure~\ref{fig:config}). These cells are supposed to be separated. In addition, let $\rmI$ and $\rmJ$ be \textit{time-intervals} with the duration $2\Ht$ on the time axis. Here, $t\ge s$ is assumed for any $t\in\rmI$ and $s\in\rmJ$. Two pairs $\OI$ and $\SJ$ are called observation and source \textit{clusters}, respectively.

To take this configuration into account, the discretised scalar and vector potentials in (\ref{eq:layer_disc}) are adjusted. First of all, the vanishing property in (\ref{eq:vanish}) is no longer valid for the present group-wise calculation. This is because the value of $\gamma^*$ in (\ref{eq:gamma*}) is point-wise and thus cannot be constant in a certain source time-interval. This situation is the same as the acoustic case described in Remark 6 of \cite{takahashi2022}, where the matrices $\mat{U}^{(\gamma)}$ and $\mat{W}^{(\gamma)}$ should be read as the matrix consisting of $\chi_{ij}^{(\gamma)}$ (for any $i,j\in[1,\Ne]$) mentioned in (\ref{eq:vanish}). To remove the vanishing property from (\ref{eq:layer_disc}), $\beta^*$ is considered as zero, that is, the summation $\sum_{\beta=\beta^*+1}^\alpha$ is replaced with $\sum_{\beta=0}^{\alpha-1}$, where the index $\beta$ is shifted by one. In addition, the resulting double-summations $\sum_{j=1}^{\Ne} \sum_{\beta=0}^{\alpha-1}$ is replaced with $\sum_{\{j~|~S_j\subset S\}} \sum_{\{\beta~|~t_\beta\in J\}}$, which will be simplified to $\sum_j \sum_\beta$. Moreover, $(\bm{x},t_\alpha)$ is supposed in $\OI$. Therefore, the following potentials are targeted to derive the fundamental three steps:
\begin{subequations}
  \begin{eqnarray}
    &&\ddot{\bm{A}}(\bm{x},t_\alpha)
    =\sum_j\sum_\beta\frac{\mu}{4\pi}\sum_{\sigma=\pm}\frac{\sigma\Ell_j}{2A_j^\sigma (c\Dt)^d}\int_{S_j^\sigma} \ddot{U}^d(\bm{x},\bm{y},t_\alpha,t_\beta)(\bm{y}-\bm{o}_j^\sigma)\ \diff S_y \hat{J}_j^\beta, \label{eq:ddotA_tmp}\\
    &&\dot{\phi}(\bm{x},t_\alpha)
    =\sum_j\sum_\beta\frac{-1}{4\pi\varepsilon}\sum_{\sigma=\pm}\frac{\sigma\Ell_j}{A_j^\sigma (c\Dt)^d}\int_{S_j^\sigma} U^d(\bm{x},\bm{y},t_\alpha,t_\beta)\ \diff S_y \hat{J}_j^\beta, \label{eq:dotphi_tmp}\\
    &&\dot{\bm{P}}(\bm{x},t_\alpha)+\zeta\bm{P}(\bm{x},t_\alpha)
    =\sum_j\sum_\beta\frac{1}{4\pi} \sum_{\sigma=\pm} \frac{\sigma\Ell_j}{2A_j^\sigma(c\Dt)^d} \int_{S_j^\sigma} (\bm{y}-\bm{o}_j^\sigma) \times\nabla_y \left(\dot{U}^d(\bm{x},\bm{y},t_\alpha,t_{\beta+\kappa})+\zeta U^d(\bm{x},\bm{y},t_\alpha,t_{\beta+\kappa})\right)\ \diff S_y \hat{J}_j^\beta,\nonumber\\
    && \label{eq:P_tmp}
  \end{eqnarray}
  \label{eq:potentials_tmp}
\end{subequations}
where $\bm{x}\in\rmO$ and $t_\alpha\in\rmI$.

\subsection{Interpolation of $U$}\label{s:fmm_expand_U}

To realise the aforementioned operations (I)--(III), it is necessary to re-express the scalar/vector potentials in (\ref{eq:potentials_tmp}) into a form of separation of variables. To this end, the IFMM interpolates the kernel function $U^d(\bm{x},\bm{y},t,s)$ in (\ref{eq:potentials_tmp}), which is defined by (\ref{eq:U}), with respect to all the eight arguments, i.e. $\bm{x}=(x_1,x_2,x_3)\in\rmO$, $\bm{y}=(y_1,y_2,y_3)\in\rmS$, $t\in\rmI$, and $s\in\rmJ$.

Regarding the temporal variable $s$, for example, let $\{\Lag_i(\cdot)\}_{n=0}^{\Pt-1}$ be a set of $\Pt$ interpolation functions (interpolants), whose domain of definition is supposed to be $[-1,1]$ and $\{\omega_n\}_{n=0}^{\Pt-1}$ be $\Pt$ interpolation points on $[-1,1]$. Then, $U^d$ can be interpolated with respect to $s\in\rmJ$ as follows:
\begin{eqnarray*}
  U^d(\bm{x},\bm{y},t,s)\approx\sum_{n=0}^{\Pt-1}U^d\left(\bm{x},\bm{y},t,\cJ+\Ht\omega_n^\Pt\right)\Lag_n\left(\frac{s-\cJ}{\Ht}\right),
\end{eqnarray*}
where $\cJ$ denotes the centre of $\rmJ$; similarly, $\cS$, $\cI$, and $\cJ$ are defined. Further, by repeating such an interpolation for the remaining seven variables (i.e. $x_1$, $x_2$, $x_3$, $y_1$, $y_2$, $y_3$, and $t$), $U^d$ can be expressed as follows:
\begin{eqnarray}
  U^d(\bm{x},\bm{y},t,s)\approx\sum_{a<\Ps}\sum_{b<\Ps}\sum_{m<\Pt}\sum_{n<\Pt}U^d_{a,b,m,n}(\rmO,\rmS,\rmI,\rmJ)
  \Lag_a\left(\frac{\bm{x}-\cO}{\Hs}\right)
  \Lag_b\left(\frac{\bm{y}-\cS}{\Hs}\right)
  \Lag_m\left(\frac{t-\cI}{\Ht}\right)
  \Lag_n\left(\frac{s-\cJ}{\Ht}\right),
  \label{eq:U_expand}
\end{eqnarray}
where $U^d_{a,b,m,n}$ denotes the value of $U^d$ at interpolation points, i.e.
\begin{eqnarray*}
  U^d_{a,b,m,n}(\rmO,\rmS,\rmI,\rmJ):=U^d(\cO+\Hs\bm{\omega}_a^\Ps,\ \cS+\Hs\bm{\omega}_b^\Ps,\ \cI+\Ht\omega_m^\Pt,\ \cJ+\Ht\omega_n^\Pt).
\end{eqnarray*}
Also, the notations
\begin{eqnarray*}
  \sum_{a<\Ps}:=\sum_{a_1=0}^{\Ps-1}\sum_{a_2=0}^{\Ps-1}\sum_{a_3=0}^{\Ps-1},\quad
  \Lag_a(\bm{\xi}):=\Lag_{a_1}(\xi_1)\Lag_{a_2}(\xi_2)\Lag_{a_3}(\xi_3),\quad
  \bm{\omega}_a^\Ps:=\left(\omega_{a_1}^\Ps, \omega_{a_2}^\Ps, \omega_{a_3}^\Ps\right)
\end{eqnarray*}
are used for the spatial index $a$. Here, $\xi_i$ is supposed to be in $[-1,1]$ that is the domain of definition of the interpolant $\Lag_{a_i}$. Similar notations are used for the index $b$.

It is found that the RHS of (\ref{eq:U_expand}) is in a form of separation variables. The expression corresponds to the multipole-expansion in the ordinary FMMs~\cite{greengard1987,liu2009fast} and the PWTD~\cite{ergin1998,ergin1999c,ergin2000,shanker2000,shanker2003,aygun2004}.

Selecting the interpolant for each variable is not an evident task from the viewpoint of the resulting accuracy. Following the previous study~\cite{takahashi2014,takahashi2022}, the present study exploits the cubic Hermite interpolation (CHI) with using finite-difference approximated derivatives for both the spatial and temporal interpolations; see \cite[Appendix~B]{takahashi2014}.

\subsection{Formulation of P2M, M2L, and L2P for the CFIE}\label{s:fmm_cfie}

The underlying formulation can be obtained by substituting the interpolated $U^d$ in (\ref{eq:U_expand}) into the potentials in (\ref{eq:potentials_tmp}). First, $\ddot{\bm{A}}$ in (\ref{eq:ddotA_tmp}) is written as follows:
\begin{eqnarray*}
  \ddot{\bm{A}}(\bm{x},t_\alpha)
  &=&\sum_j\sum_\beta\frac{\mu}{4\pi}\sum_{\sigma=\pm}\frac{\sigma\Ell_j}{2A_j^\sigma (c\Dt)^d}\int_{S_j^\sigma} \sum_a\sum_b\sum_m\sum_n \dot{U}^d_{a,b,m,n}(\rmO,\rmS,\rmI,\rmJ)\nonumber\\
  &&\Lag_a\left(\frac{\bm{x}-\cO}{\Hs}\right) \Lag_b\left(\frac{\bm{y}-\cS}{\Hs}\right) \dot{\Lag}_m\left(\frac{t_\alpha-\cI}{\Ht}\right) \Lag_n\left(\frac{t_\beta-\cJ}{\Ht}\right)(\bm{y}-\bm{o}_j^\sigma)\diff S_y \hat{J}_j^\beta.
\end{eqnarray*}
To derive this, first, the substitution of $t=t_\alpha$ is withdrawn in (\ref{eq:ddotA_tmp}). Second, one temporal-differentiation $\partial_t$ is detached from $\ddot{U}^d$. Third, the first-order derivative $\dot{U}^d$ ($\equiv cd U^{d-1}$; recall Eq.(\ref{eq:U})) is interpolated in the same way as (\ref{eq:U_expand}). Fourth, the generated interpolant $\Lag_m((t-\cI)/\Ht)$ is differentiated with the detached $\partial_t$. Finally, $t_\alpha$ is plugged into $t$. On contrary, it is possible to apply $\partial_t^2$ to either $U^d$ or $\Lag_m$, but separating two $\partial_t$s seems the best; this will be discussed in Section~\ref{s:test_discuss}. 

Next, arranging the order of the summations reduces the above expression to
\begin{eqnarray}
  \ddot{\bm{A}}(\bm{x},t_\alpha)
  &=&\sum_a\sum_m \Lag_a\left(\frac{\bm{x}-\cO}{\Hs}\right) \dot{\Lag}_m\left(\frac{t_\alpha-\cI}{\Ht}\right) \sum_b \sum_n \dot{U}^d_{a,b,m,n}(\rmO,\rmS,\rmI,\rmJ)\nonumber\\
  &&\underbrace{\sum_j\sum_\beta \frac{\mu}{4\pi}\sum_{\sigma=\pm}\frac{\sigma\Ell_j}{2A_j^\sigma (c\Dt)^d}\int_{S_j^\sigma} \Lag_b\left(\frac{\bm{y}-\cS}{\Hs}\right) (\bm{y}-\bm{o}_j^\sigma)\diff S_y\Lag_n\left(\frac{t_\beta-\cJ}{\Ht}\right) \hat{J}_j^\beta }_{\displaystyle=:\bm{M}_{b,n}(\rmS,\rmJ)},
  \label{eq:ddotA_M}
\end{eqnarray}
where $\bm{M}_{b,n}$ corresponds to the MM in the terminology of the FMM. The definition of the MM corresponds to the P2M for $\ddot{A}$. It should be noted that $\bm{M}_{b,n}$ is a three-dimensional vector. Also, the integral over $\bm{y}$ in $\bm{M}_{b,n}$ (as well as the following MMs) is evaluated with the Gaussian quadrature formula; this study uses three-point formula.

Successively, the corresponding local-coefficient $\bm{L}_{a,m}$ can be defined as follows:
\begin{eqnarray}
  \ddot{\bm{A}}(\bm{x},t_\alpha)
  =\sum_a\sum_m \Lag_a\left(\frac{\bm{x}-\cO}{\Hs}\right) \dot{\Lag}_m\left(\frac{t_\alpha-\cI}{\Ht}\right) \underbrace{\sum_b \sum_n \dot{U}^d_{a,b,m,n}(\rmO,\rmS,\rmI,\rmJ) \bm{M}_{b,n}(\rmS,\rmJ)}_{\displaystyle=:\bm{L}_{a,m}(\rmO,\rmI)}.
  \label{eq:ddotA_fmm}
\end{eqnarray}
The above definition of $\bm{L}_{a,m}$ is nothing but the M2L formula. In addition, the above equation designates the L2P for the vector potential $\ddot{A}$.

Second, similarly to $\ddot{A}$, one can derive the fundamental FMM-operations for the time-differentiated scalar potential $\dot{\phi}$ from (\ref{eq:dotphi_tmp}) as follows:
\begin{eqnarray}
  \dot{\phi}(\bm{x},t_\alpha)
  &=& \sum_a\sum_m \Lag_a\left(\frac{\bm{x}-\cO}{\Hs}\right) \Lag_m\left(\frac{t_\alpha-\cI}{\Ht}\right) \sum_b\sum_n U^d_{a,b,m,n}(\rmO,\rmS,\rmI,\rmJ)\nonumber\\
  && \underbrace{\sum_j\sum_\beta \frac{-1}{4\pi\varepsilon}\sum_{\sigma=\pm} \frac{\sigma\Ell_j}{A_j^\sigma(c\Dt)^d}\int_{S_j^\sigma} \Lag_b\left(\frac{\bm{y}-\cS}{\Hs}\right) \diff S_y \Lag_n\left(\frac{t_\beta-\cJ}{\Ht}\right) \hat{J}_j^\beta}_{\displaystyle=:M_{b,n}(\rmS,\rmJ)}\nonumber\\
  &=& \sum_a\sum_m \Lag_a\left(\frac{\bm{x}-\cO}{\Hs}\right) \Lag_m\left(\frac{t_\alpha-\cI}{\Ht}\right) \underbrace{\sum_b\sum_n U^d_{a,b,m,n}(\rmO,\rmS,\rmI,\rmJ) M_{b,n}(\rmS,\rmJ)}_{\displaystyle=:L_{a,m}(\rmO,\rmI)}.
  \label{eq:dotphi_fmm}
\end{eqnarray}
In this case, both the MM $M_{b,n}$ and the LC $L_{a,m}$ are scalars.

Third and last, from (\ref{eq:P_tmp}), the vector potential $\dot{\bm{P}}+\zeta\bm{P}$ can be represented as follows:
\begin{eqnarray}
  \dot{\bm{P}}(\bm{x},t_\alpha)+\zeta\bm{P}(\bm{x},t_\alpha)
  &=&\sum_a\sum_m\Lag_a\left(\frac{\bm{x}-\cO}{\Hs}\right) \left( \dot{\Lag}_m\left(\frac{t_\alpha-\cI}{\Ht}\right)+\zeta\Lag_m\left(\frac{t_\alpha-\cI}{\Ht}\right)\right) \sum_b\sum_n U^d_{a,b,m,n}(\rmO,\rmS,\rmI,\rmJ)\nonumber\\
  &&\underbrace{\sum_j\sum_\beta\sum_{\sigma=\pm} \frac{1}{4\pi}\frac{\sigma\Ell_j}{2A_j^\sigma(c\Dt)^d}\int_{S_j^\sigma} (\bm{y}-\bm{o}_j^\sigma)\times\nabla_y\Lag_b\left(\frac{\bm{y}-\cS}{\Hs}\right)\diff S_y\Lag_n\left(\frac{t_\beta-\cJ}{\Ht}\right) \hat{J}_j^\beta}_{\displaystyle=:\bm{M}^{\mathrm{mag}}_{b,n}(\rmS,\rmJ)}\nonumber\\
  &=&\sum_a\sum_m \Lag_a\left(\frac{\bm{x}-\cO}{\Hs}\right) \left( \dot{\Lag}_m\left(\frac{t_\alpha-\cI}{\Ht}\right)+\zeta\Lag_m\left(\frac{t_\alpha-\cI}{\Ht}\right)\right) \underbrace{\sum_b\sum_n U^d_{a,b,m,n}(\rmO,\rmS,\rmI,\rmJ) \bm{M}^{\mathrm{mag}}_{b,n}(\rmS,\rmJ)}_{\displaystyle=:\bm{L}_{a,m}^{\mathrm{mag}}(\rmO,\rmI)}.\nonumber\\
  &&%
  \label{eq:P_fmm}
\end{eqnarray}
Here, another vectorial MM and LC, denoted by $\bm{M}_{b,n}^{\mathrm{mag}}$ and $\bm{L}_{a,m}^{\mathrm{mag}}$, respectively, are defined. Also, the temporal differentiation of $\dot{\bm{P}}$ is not applied to $U^d$ but $\Lag_m$ in (\ref{eq:P_fmm}). In this case, the number and type of MMs and LCs do not change from those when using either the MFIE or $\partial_t$MFIE. On the other hand, in the case that the temporal differentiation is applied to $U^d$, another vectorial MM and LC associated with the first-order derivative $\dot{U}^d$ are necessary in addition to those associated with $U^d$. This case will not be considered to save the computation time and memory in this study.

From (\ref{eq:ddotA_fmm}), (\ref{eq:dotphi_fmm}), and (\ref{eq:P_fmm}), the $\DtEFIE$ and $\DtMFIE+\zeta\MFIE$ tested by the RWG function $\bm{f}_i$, whose centre is included in $\rmO$, can be represented at $t=t_\alpha$ as follows:
\begin{subequations}
  \begin{eqnarray}
    &&\int_S\bm{f}_i(\bm{x})\cdot\left[-\bm{n}(\bm{x})\times\bm{n}(\bm{x})\times\left(\ddot{\bm{A}}(\bm{x},t_\alpha)+\nabla_{S_x}\dot{\phi}(\bm{x},t_\alpha)\right)\right]\diff S_x\nonumber\\
    &=&\sum_{\rho=\pm}\int_{S_i^\rho}\sum_a\sum_m \Lag_a\left(\frac{\bm{x}-\cO}{\Hs}\right)  \left[
      \dot{\Lag}_m\left(\frac{t_\alpha-\cI}{\Ht}\right) \bm{f}_i(\bm{x})\cdot\bm{L}_{a,m}(\rmO,\rmI)+\Lag_m\left(\frac{t_\alpha-\cI}{\Ht}\right) (\nabla_{S_x}\cdot\bm{f}_i(\bm{x})) L_{a,m}(\rmO,\rmI)
      \right]\diff S_x,\\
    &&\int_{S} \bm{f}_i(\bm{x}) \cdot \left[\bm{n}(\bm{x})\times\left(\dot{\bm{P}}(\bm{x},t_\alpha)+\zeta\bm{P}(\bm{x},t_\alpha)\right)\right] \diff S_x\nonumber\\
    &=&\sum_{\rho=\pm}\int_{S_i^\rho} \bm{f}_i(\bm{x}) \cdot  \left[ \bm{n}(\bm{x})\times 
    \sum_a\sum_m \Lag_a\left(\frac{\bm{x}-\cO}{\Hs}\right) \left( \dot{\Lag}_m\left(\frac{t_\alpha-\cI}{\Ht}\right)+\zeta\Lag_m\left(\frac{t_\alpha-\cI}{\Ht}\right)\right) \bm{L}_{a,m}^{\mathrm{mag}}(\rmO,\rmI)\right]\diff S_x.
  \end{eqnarray}
  \label{eq:eval}%
\end{subequations}%
These expressions are used to compute the far-field components of $\sum_{\beta=\beta^*+1}^\alpha\mat{Z}^{(\alpha-\beta+1)}\hat{\mat{J}}^{\beta-1}$ in (\ref{eq:R}). The far-field computation is neither RWG basis-wise nor time-step-wise on contrary to the remaining computation i.e. the near-field computation.

\def\refalgo#1{Program~\ref{#1}}
\def\refalgos#1{Programs~\ref{#1}}
\def\gammamin{\gamma_{\mathrm{min}}}
\def\acc{+=}

\iftrue
\def\STATE{\State}
\def\FOR{\For}
\def\ENDFOR{\EndFor}
\def\IF{\If}
\def\ELSE{\Else}
\def\ENDIF{\EndIf}
\def\COMMENT{\Comment}
\fi

\subsection{Algorithm of the fast TDBEM}\label{s:fmm_algo}

The overall algorithm of the fast TDBEM for electromagnetics are almost the same as those for acoustics~\cite{takahashi2022}. Simply speaking, the EM case handles multiple and vectorial MMs and LCs, whereas the acoustic case does single and scalar ones. Hence, an operation regarding the single and scalar MM/LC of the acoustic IFMM may be performed seven times, i.e. one scalar MM/LC and six components of the two vectorial MMs/LCs.

\subsubsection{Space-time hierarchy}\label{s:hierarchy}

A spatial hierarchy of $\Ne$ edges (or RWG bases) involved in the boundary $S$ is generated in the following way. First, $S$ is surrounded by a cubic box, called the root cell at \textit{level} 0. Next, the \textit{root} cell is subdivided to eight sub-cubes, called cells at level 1. Here, cells with no edge are discarded. Successively, a subdivision is repeated to every cell, say $C$, at a certain level $l$ to generate its \textit{child cells} at level $l+1$ if $C$ contains a predefined number $\Nleaf$ or more number of edges. A cell with no child cell is called a \textit{leaf cell}. As a result, an \textit{adaptive octree} whose maximum level is denoted by $\maxlev$ is created. The size of cells at level $l$ is denoted by $\ds^{(l)}$ or $2\Hs^{(l)}$, where $\ds^{(l)}\equiv 2\Hs^{(l)}$ and $\Hs^{(l)}\equiv\ds^{(l+1)}$ hold. Here and hereafter, the superscript `$(l)$' of a symbol expresses that the symbol depends on the level $l$ of the octree, where $l\in[0,\maxlev]$.

Then, according to the original FMM~\cite{greengard1987}, the \textit{neighbour cells} of $C$ at a certain level are defined with at most $27$ ($=3^3$) cells that contact with $C$ at the same level. The list of neighbour cells of $C$ is denoted by $\NL(C)$. On the other hand, the \textit{interacting cells} of $C$ at a level is defined with at most $189$ ($=6^3-3^3$) cells that are (i) children of all the neighbour cells of $C$'s parent and (ii) cells that do not touch $C$. The list of interacting cells, which is the so-called interaction-list, is denoted by $\IL(C)$.

The construction of the temporal hierarchy of time-steps, i.e. $t_0,\ldots,t_{\Nt}$, is associated with the levels of the above octree. At each level $l$, the time-axis is split to time-intervals whose durations are the same and denoted by $\dt^{(l)}$ or $2\Ht^{(l)}$, where $l\in[0,\maxlev]$ and $\dt^{(l)}\equiv 2\Ht^{(l)}$ holds. Accordingly, the $k$th time-interval at level $l$ is denoted by $I_k^{(l)}$, where $k\ge 0$.

As seen in the previous section, the IFMM conveys an influence of a source cluster $\SJ$ to an observation one $\OI$, where it is assumed that $\rmS\in\IL(\rmO)$ and $\rmI$ is in the future side of $\rmJ$. Hence, if the duration $\dt^{(l)}$ is determined as the minimum of the travelling time from $\rmO$ to $\rmS$, i.e. $\dt^{(l)}:=\ds^{(l)}/c$, the source information of $\rmS$ generated during $\rmJ$ transmits to $\rmO$ after the end of $\rmJ$. Hence, when a source time-interval $\rmJ$ is $I_k^{(l)}$, then $\rmI$ can be $I_{k+1}^{(l)}$, $I_{k+2}^{(l)}$, $\ldots$. More precisely, the number of time-steps in each time-interval at the maximum level $\maxlev$ is first determined by
\begin{eqnarray}
  \Mt^{(\maxlev)}:=\mathrm{ceil}\left(\frac{\ds^{(\maxlev)}}{c\Dt}\right).
  \label{eq:Mt}
\end{eqnarray}
Then, the corresponding duration is determined by
\begin{eqnarray*}
  \dt^{(\maxlev)}:=\Mt^{(\maxlev)}\Dt.
\end{eqnarray*}
The number of time-steps and duration at level $l$ is recursively determined as follows:
\begin{eqnarray*}
  \Mt^{(l)}:=2\Mt^{(l+1)},\quad \dt^{(l)}:=2\dt^{(l+1)}\quad l=\maxlev-1,\ldots,2.
\end{eqnarray*}
By construction, the children of $I_k^{(l)}$ correspond to $I_{2k}^{(l+1)}$ and $I_{2k+1}^{(l+1)}$.

\begin{remark}\label{remark:breakdown}
When $\Dt$ is large or $\ds^{(\maxlev)}$ is small, the number $\Mt^{(\maxlev)}$ in (\ref{eq:Mt}) can become zero and, thus, the present algorithm breaks down.
\end{remark}

\subsubsection{Description of the fast algorithm with pseudo programs}\label{s:pseudo}

\refalgo{algo:main} is the main program to solve the discretised CFIE in (\ref{eq:solve}) in the MOT scheme. After building the aforementioned space-time hierarchy from a part of the inputs (i.e. mesh for $S$, $\Dt$, $\Nt$, and $\Nleaf$), all the RHS vectors $\mat{R}^0,\ldots,\mat{R}^{\Nt-2}$ are initialised at Line~\ref{line:main_init}. This is because the influence of the current time-interval is thrown to some RHS vectors of future time-intervals at Line~\ref{line:main_far}. In accordance with the principle of the FMM, the present IFMM conceptually decomposes each coefficient matrix in (\ref{eq:R}), say $\mat{Z}^{(\gamma)}$, to its near- and far-field components, say $\mat{Z}^{(\gamma)}_{\rm near}$ and $\mat{Z}^{(\gamma)}_{\rm far}$, respectively. The former $\mat{Z}^{(\gamma)}_{\rm near}$ consists of the interactions among RWG bases when they are close to one another. The closeness is judged in the grain of cells: Two cells $\rmO$ and $\rmS$ at the same level are close to each other if $\rmS\in\NL(\rmO)$ ($\Leftrightarrow \rmO\in\NL(\rmS)$). The matrix $\mat{Z}^{(\gamma)}_{\rm near}$ thus computed is multiplied to a certain vector $\hat{\mat{J}}^{\beta-1}$; in practise, the matrix-vector product is performed on the fly without allocating the matrix $\mat{Z}^{(\gamma)}_{\rm near}$ in computer memory. The near-field contribution thus obtained is accumulated to the current RHS vector $\mat{R}^{\alpha-1}$; see Line~\ref{line:main_near}. More details are described in \refalgo{algo:near}. The near-field computation is called P2P or direct computation in the FMM terminology.

\begin{algorithm}[H]
  \caption{Main.}
  \label{algo:main}
  \begin{algorithmic}[1]
    \STATE Input: mesh for $S$ (consisting of $\Ns$ triangles), permittivity $\epsilon$, permeability $\mu$, incident fields $\Einc$ and $\Hinc$, coupling parameters $\theta$ and $\zeta$, number $\Nt$ of time-steps,  time-step size $\Dt$, numbers $\Ps$ and $\Pt$ of interpolation pints, maximum number $\Nleaf$ of edges in a leaf cell (recall Section~\ref{s:hierarchy}), number of Gaussian quadrature points (recall Section~\ref{s:fmm_cfie}), parameters for the GMRES (recall Section~\ref{s:tdbem_solve}).
    
    \STATE Build the space-time hierarchy according to Section~\ref{s:hierarchy}.
    
    \STATE Initialise all the RHS vectors $\mat{R}^{\alpha-1}=\mat{0}$ for $\alpha=1,\ldots,\Nt-1$. \label{line:main_init}
    
    \FOR{time-step $\alpha=1$ to $\Nt-1$}
    
    \STATE Add the incident field $\mat{V}^{\alpha-1}$ to $\mat{R}^{\alpha-1}$.
    
    \STATE Add the near-field contribution from the current time-step $\alpha$ to $\mat{R}^{\alpha-1}$ according to \refalgo{algo:near}. \label{line:main_near}
    
    \STATE Solve (\ref{eq:solve}) for the unknown vector $\mat{J}^{\alpha-1}$ with using the GMRES. \label{line:main_solve}
    
    \STATE Add the far-field contribution from the current time-interval to $\mat{R}^\beta$ regarding a finite number of future time-intervals according to \refalgos{algo:upward} and \ref{algo:downward}. \label{line:main_far}
    
    \ENDFOR
  \end{algorithmic}
\end{algorithm}

In \refalgo{algo:near}, the definition of the number $\gamma^{*(l)}$ at Line~\ref{line:near_gamma} is similar to that of $\gamma^*$ in (\ref{eq:gamma*}), but the domain of $\bm{x}$ and $\bm{y}$ is limited to $\rmO$'s neighbours, i.e.
\begin{eqnarray}
  \gamma^{*(l)}:=\frac{\mathrm{max}_{\bm{x}\in\rmO,\ \bm{y}\in\NL(\rmO)}|\bm{x}-\bm{y}|}{c\Dt}=\frac{2\sqrt{3}\ds^{(l)}}{c\Dt}.
\end{eqnarray}
From this, as the level $l$ increases (that is, the spatial scale becomes smaller), the number of the \textit{passed} time-steps, which corresponds to the loop over $\beta$ at Line~\ref{line:near_gamma}, decreases and, thus, the amount of direct computation decreases. Hence, the near-field computation is localised in both space and time. 

\begin{algorithm}[H]
  \caption{Near-field computation (P2P).}
  \label{algo:near}
  \begin{algorithmic}[1]
    \STATE{The current time-step is $\alpha$.}

    \FOR{level $l=2$ to $\maxlev$}\label{line:near_level}

    \FOR{cell $\rmO$ in level $l$}\label{line:near_O}

    \FOR{cell $\rmS$ in the neighbour-list $\NL(\rmO)$}
    
    \IF{$\rmO$ or $\rmS$ is a leaf cell}
    
    \FOR{RWG basis $\bm{f}_i$ in $\rmO$} \label{line:near_fi}
    
    \FOR{RWG basis $\bm{f}_j$ in $\rmS$} \label{line:near_fj}
    
    \FOR{time-step $\beta=\alpha-\gamma^{*(l)}+1$ to $\alpha$} \label{line:near_gamma}
    
    \STATE Calculate and store the coefficient $Z_{ij}^{(\alpha-\beta+1)}$ if it has not stored yet.

    \STATE Add the product $-Z_{ij}^{(\alpha-\beta+1)}\hat{J}^{\beta-1}_j$ to $R_i^{\alpha-1}$.

    \ENDFOR
    \ENDFOR
    \ENDFOR
    \ENDIF
    \ENDFOR
    \ENDFOR
    \ENDFOR
  \end{algorithmic}
\end{algorithm}

After solving (\ref{eq:solve}) for the current unknown vector $\mat{J}^{\alpha-1}$ (Line~\ref{line:main_far} of \refalgo{algo:main}), the far-field contribution are computed at Line~\ref{line:main_far} of \refalgo{algo:main} through the upward and downward passes in Programs~\ref{algo:upward} and \ref{algo:downward}, respectively. These passes are performed when the current time-step $\alpha$ coincides with the end of the current time-interval $I_k^{(l)}$, where $k:=\alpha/\Mt^{(l)}-1$, at level $l$. (This is called \textit{time-gating} according to the PWTD.) Then, the far-field computation can be explained with the M2L operation, by which the far-field contribution from a cluster $(O, I_k^{(l)})$ is thrown to any clusters $(S, I_{j}^{(l)})$ such that the interacting cells $S\in\IL(O)$ and the future time-intervals $j=k+1,\ldots,k+\mu+1$. Here, $\mu$ is determined as $8$ regardless of levels~\cite[Section~4.2]{takahashi2022}. As a result, the future RHS vectors (i.e. $\mat{R}^{\alpha},\mat{R}^{\alpha+1}, \ldots$) that are related to the future $\mu+2$ time-intervals are updated at Line~\ref{line:main_far} of \refalgo{algo:main}.

It should be noted that the M2L translation in \refalgo{algo:downward} is not based on that in Code~4 of \cite{takahashi2014} but Algorithm~1 of \cite{takahashi2022}. This is because Algorithm~1 (or Formula~1) of \cite{takahashi2022} is the generalisation of Code~4 of \cite{takahashi2014} from $d=1$ to $d\ge 1$, where $d$ denotes the order of the B-spline temporal basis; recall the approximation of $\bm{J}$ in (\ref{eq:J}).

The present upward and downward passes are similar to those of the acoustic case, i.e. Codes~3 and 4 of \cite{takahashi2014}, but different in the following aspects:
\begin{itemize}

\item Each of the M2M (Line~\ref{line:upward_m2m} of \refalgo{algo:upward}), M2L (Line~\ref{line:downward_m2l} of \refalgo{algo:downward}), and L2L (Line~\ref{line:downward_l2l} of \refalgo{algo:downward}) is performed not once but seven times because the present EM case considers one scalar and two vectorial MMs and LCs on contrary to one scalar MM and LC in the acoustic case.

\item As shown in Line~\ref{line:downward_l2p} of \refalgo{algo:downward}, the scalar and vector potentials are integrated on triangles according to (\ref{eq:eval}) in the EM case, while the layer-potential is evaluated at collocation points in the acoustic case. However, since the integrals over $\bm{x}$ in (\ref{eq:eval}) are evaluated by the Gaussian quadrature formula, the quadrature points can be regarded as the collocation points of the acoustic case. Hence, the existence of the integrals over $\bm{x}$ is irrelevant to the concept of the IFMM.

\end{itemize}
From these, the EM case actually requires more computation time and memory than the acoustic case, although the computation complexity of the former is the same as that of the latter (see Section~\ref{s:fmm_complexity}). Since the most of computation time is spent for the M2L translation in the present multi-level algorithm, the computation time of the EM case approximately increases by a factor of $7 \times 3$, where the number `$7$' represents the ratio of the number of MMs/LCs for the EM case to that for the acoustic case and `$3$' represents the number of the Gaussian quadrature points for the integrals over $\bm{x}$.

\begin{algorithm}[H]
\caption{Upward pass for the far-field computation.}
  \label{algo:upward}
  \begin{algorithmic}[1]
    \STATE The current time-step is $\alpha$.

    \FOR{level $l=\maxlev$ to $2$}
   
    \IF{$\alpha$ is divisible by $\Mt^{(l)}$}
    
    \STATE The current time-interval is $I_k^{(l)}$, where $k:=\frac{\alpha}{\Mt^{(l)}}-1$.

    \FOR{cell $\rmS$ in level $l$}\label{line:upward_S}
    
    \IF{$\rmS$ is a leaf cell}
    
    \STATE Perform the P2M operation according to (\ref{eq:ddotA_M}), (\ref{eq:dotphi_fmm}), and (\ref{eq:P_fmm}). \label{line:upward_p2m}
    
    \ELSE
    
    \STATE Perform the M2M operation; compute each of the seven MMs of $(\rmS,\rmI_k^{(l)})$ from those of $(\rmS',\rmI_{2k}^{(l+1)})$ and $(\rmS',\rmI_{2k+1}^{(l+1)})$ for every $\rmS$'s child cell $\rmS'$ according to the M2M formula in (16) of \cite{takahashi2014}. \label{line:upward_m2m}

    \ENDIF
    \ENDFOR
    \ENDIF
    \ENDFOR
  \end{algorithmic}
\end{algorithm}

\begin{algorithm}[H]
  \caption{Downward pass for the far-field computation.}
  \label{algo:downward}
  \begin{algorithmic}[1]
    \STATE The current time-step is $\alpha$.

    \FOR{level $l=2$ to $\maxlev$}

    \IF{$\alpha$ is divisible by $\Mt^{(l)}$}

    \STATE The current time-interval is $I_k^{(l)}$, where $k:=\frac{\alpha}{\Mt^{(l)}}-1$.   

    \FOR{cell $\rmO$ in level $l$}\label{line:downward_O}

    \STATE Perform the M2L operation for each of seven pairs of MM and LC in (\ref{eq:ddotA_fmm}), (\ref{eq:dotphi_fmm}), and (\ref{eq:P_fmm}). Specifically, Algorithm~4 in \cite{takahashi2022} is applied to every pair. \label{line:downward_m2l}

    \IF{$\rmO$ is a leaf cell} 

    \STATE Compute the integrals for any $\bm{f}_i\in\rmO$ and $t_\beta\in\rmI_{k+1}^{(l)}$ with the LCs of $(\rmO,\rmI_{k+1}^{(l)})$ according to (\ref{eq:eval}), where $\rmI$ and $t_\alpha$ read $\rmI_{k+1}^{(l)}$ and $t_\beta$, respectively. \label{line:downward_l2p}

    \ELSE
    
    \STATE Perform the L2L operation; compute each of the seven LCs of $(\rmO',\rmI_{2(k+1)}^{(l+1)})$ and $(\rmO',\rmI_{2(k+1)+1}^{(l+1)})$ from that of $(\rmO,\rmI_{k+1}^{(l)})$ for every $\rmO$'s child $\rmO'$ according to (17) of \cite{takahashi2014}. \label{line:downward_l2l}
    
    \ENDIF
    
    \ENDFOR
    \ENDIF
    \ENDFOR
  \end{algorithmic}
\end{algorithm}

\subsection{Complexity}\label{s:fmm_complexity}

In addition to Assumptions~\ref{assume:S} and \ref{assume:Dt}, the following assumptions are considered in order to compute the complexity of the fast TDBEM:
\begin{assumption}\label{assume:nleaf}
  The parameters $\Nleaf$, $\Ps$, and $\Pt$ of the IFMM are independent of $\Ns$ and $\Nt$, i.e. $\Nleaf=\Order(1)$,  $\Ps=\Order(1)$, and $\Pt=\Order(1)$.
\end{assumption}
\begin{assumption}\label{assume:48}
  \begin{subequations}
    The $\Ne$ edges (or RWG bases) are distributed uniformly in the cubic domain (or root cell) so that each non-leaf cell can have fully eight children. Moreover, every leaf cell exists at level $\maxlev$. Then, it follows that
    \begin{eqnarray}
      \Nleaf\cdot 8^\maxlev=\Ne\quad\Longleftrightarrow\quad 2^\maxlev=\left(\frac{\Ne}{\Nleaf}\right)^{1/3}\sim\Order(\Ns^{1/3}).
      \label{eq:assume8}
    \end{eqnarray}
    Alternatively, $\Ne$ edges are distributed on a plane in the cubic domain so that each non-leaf cell can have four children. Moreover, every leaf cell exists at level $\maxlev$. These lead to
    \begin{eqnarray}
      \Nleaf\cdot 4^\maxlev=\Ne\quad\Longleftrightarrow\quad 2^\maxlev=\left(\frac{\Ne}{\Nleaf}\right)^{1/2}\sim\Order(\Ns^{1/2}).
      \label{eq:assume4}
    \end{eqnarray}%
  \end{subequations}%
\end{assumption}

\begin{remark}\label{remark:complexity}
  The computational complexity of the proposed fast TDBEM is estimated as $\Order(\Ns^{4/3}\Nt)$ (respectively, $\Order(\Ns^{3/2}\Nt)$) under the assumptions in Assumptions \ref{assume:S}, \ref{assume:Dt}, \ref{assume:nleaf}, and (\ref{eq:assume8}) (respectively, (\ref{eq:assume4})) of Assumption~\ref{assume:48}.
\end{remark}

For clarity, the computational complexity in the uniform-distribution case of (\ref{eq:assume8}) is described similarly to \cite[Appendix~A]{takahashi2014}. To this end, the following relationships are available:
\begin{eqnarray*}
  \Dt,\ \Nleaf,\ \Ps,\ \Pt\sim\Order(1),\quad
  \Ds^{(l)}\sim 2^{-l},\quad
  \gamma^{*(l)}\sim\frac{\Ds^{(l)}}{\Dt}\sim 2^{-l}\cdot\Dt^{-1},\quad
  2^{\maxlev}\sim\Ns^{1/3},\quad
  \Mt^{(l)}\sim\frac{\Ds^{(l)}}{\Dt}\sim 2^{-l}\cdot\Dt^{-1}.
\end{eqnarray*}
Here and in what follows, a constant $\Dt$ is not dropped in order to consider another scenario of the computational complexity, which will be mentioned in Remark~\ref{remark:fix_Dt}. From these, the floating-operation counts (flops) of major operations in the IFMM are calculated as follows:
\begin{enumerate}
\item P2P at Line~\ref{line:main_near} in \refalgo{algo:main}:

  \begin{eqnarray*}
    &&(\mbox{flops per leaf and time-step})
    \times(\mbox{\# of leaves at level $\maxlev$})
    \times(\mbox{\# of time-steps})\\
    &\sim& \Nleaf \cdot 27 \Nleaf \cdot \gamma^{*(\maxlev)} \times 8^{\maxlev} \times \Nt\\
    &\sim& 27\Nleaf^2 \cdot 2^{-\maxlev}\cdot\Dt^{-1} \times \frac{\Ns}{\Nleaf} \times \Nt
    \sim \Ns^{2/3} \Nt \Dt^{-1}
  \end{eqnarray*}

\item P2M at Line~\ref{line:upward_p2m} in \refalgo{algo:upward}:

  \begin{eqnarray*}
    &&(\mbox{flops per leaf and time-interval})
    \times(\mbox{\# of leaves at level $\maxlev$})
    \times(\mbox{\# of time-intervals})\\
    &\sim& \Nleaf \Mt^{(\maxlev)} \Ps^3\Pt \times 8^{\maxlev}\times\frac{\Nt}{\Mt^{(\maxlev)}}
    \sim \Ns\Nt
  \end{eqnarray*}

\item M2M at Line~\ref{line:upward_m2m} in \refalgo{algo:upward}:

  \begin{eqnarray*}
    &&\sum_{l=2}^{\maxlev-1}
    (\mbox{flops per cell and time-interval})
    \times(\mbox{\# of cells})
    \times(\mbox{\# of time-intervals})\\
    &\sim& \sum_{l=2}^{\maxlev-1} \Ps^6\Pt^2 \cdot 8 \times 8^l\times\frac{\Nt}{\Mt^{(l)}} 
    \sim \sum_{l=2}^{\maxlev-1} 16^l \Nt\Dt
    \sim 16^\maxlev \Nt\Dt
    \sim \Ns^{4/3}\Nt\Dt
  \end{eqnarray*}
  
\item M2L at Line~\ref{line:downward_m2l} in \refalgo{algo:downward}:

  As noted in Section~\ref{s:pseudo}, the M2L formula of \cite[Formula~1]{takahashi2022} is applied to each pair of MM and LC for the EM case. Then, not only computing the LCs according to (\ref{eq:ddotA_fmm}), (\ref{eq:dotphi_fmm}), and (\ref{eq:P_fmm}) but also computing the \textit{derivatives of LCs} as well as the \textit{associated LCs} are necessary. However, the most expensive computation is the computation of the LCs. The corresponding flops are estimated as follows:
  \begin{eqnarray*}
    &&\sum_{l=2}^{\maxlev}
    (\mbox{flops per cell and time-interval})
    \times(\mbox{\# of cells})
    \times(\mbox{\# of time-intervals})\\
    &\sim& \sum_{l=2}^{\maxlev} \Ps^3\Pt(\log\Ps+\log\Pt) \cdot 189 \cdot 9 \times 8^l \times \frac{\Nt}{\Mt^{(l)}}\\
    &\sim& \sum_{l=2}^{\maxlev}  16^l\Nt\Dt
    \sim 16^{\maxlev}\Nt\Dt
    \sim \Ns^{4/3} \Nt \Dt
  \end{eqnarray*}
  Here, the factor `$9$' in the second line denotes the number of future time-intervals to be considered; this was mentioned as `$\mu+1$' in Section~\ref{s:fmm_algo} and will be mentioned in Section~\ref{s:fmm_remark_num}.

\item L2L at Line~\ref{line:downward_l2l} in \refalgo{algo:downward}:

  Since this is almost the same operation as the M2M translation, the flops count scales as $\Ns^{4/3}\Nt\Dt$.

\item L2P at Line~\ref{line:downward_l2p} in \refalgo{algo:downward}:

  \begin{eqnarray*}
    &&(\mbox{flops per leaf and time-interval})
    \times(\mbox{\# of leaves at level $\maxlev$})
    \times(\mbox{\# of time-intervals})\\
    &\sim& \Nleaf \Mt^{(\maxlev)} \Ps^3\Pt \times 8^{\maxlev}\times\frac{\Nt}{\Mt^{(\maxlev)}}
    \sim \Ns\Nt
 \end{eqnarray*}

\end{enumerate}

From these counts, the computational complexity is indeed estimated as $\Order(\Ns^{4/3}\Nt)$ under the assumption of the uniform distribution in (\ref{eq:assume8}).

\begin{remark}\label{remark:fix_Dt}
  Instead of letting $\Dt$ be a constant as in Assumption~\ref{assume:Dt}, another scenario is to fix the final time $T$, i.e. $T:=(\Nt-1)\Dt\sim\Nt\Dt$. In this case, $\Dt$ scales as $\Order(\Nt^{-1})$. Therefore, the complexities of the operations in the fast TDBEM are calculated as 
  \begin{eqnarray*}
    \text{P2P}\sim \Ns^{2/3}\Nt^2,\quad
    \text{P2M, L2P}\sim \Ns\Nt,\quad
    \text{M2M, M2L, L2L}\sim \Ns^{4/3} 
 \end{eqnarray*}
  on the assumption of (\ref{eq:assume8}) and
  \begin{eqnarray*}
    \text{P2P}\sim \Ns^{1/2}\Nt^2,\quad
    \text{P2M, L2P}\sim \Ns\Nt,\quad
    \text{M2M, M2L, L2L}\sim \Ns^{3/2}
  \end{eqnarray*}
  on the alternative assumption of (\ref{eq:assume4}). Further, if both $\Ns$ and $\Nt$ vary in proportional to a certain parameter $n$, i.e. $\Ns,\Nt\sim\Order(n)$, the computational complexity of the fast TDBEM is calculated as $\Order(n^{8/3})$ or $\Order(n^{5/2})$ for (\ref{eq:assume8}) and (\ref{eq:assume4}), respectively. On the other hand, because $\gamma^*$ in (\ref{eq:gamma*}) scales as $\Order(\Dt)=\Order(\Nt^{-1})$, the complexity of the conventional TDBEM is estimated as $\Order(\Ns^2\Nt^2)=\Order(n^4)$. 
\end{remark}

In the numerical validation of the previous work for acoustics~\cite[Section~5.2]{takahashi2014}, a series of problems were analysed in the case of $\Dt=\Order(\Nt^{-1})$ and $\Ns$, $\Nt\sim\Order(n)$ as in Remark~\ref{remark:fix_Dt}. However, the actual computation times of the conventional and fast TDBEMs for acoustics were discussed in comparison with the computational complexities in case that $\Dt$ is fixed, which correspond to those in Section~\ref{s:tdbem_complexity} and Remark~\ref{remark:complexity}, respectively. In this regard, the discussion is incomplete, but it is true that the fast TDBEM was essentially faster than the conventional one.

\subsection{Notes}\label{s:fmm_note}

Before closing this section, some notes are made with regard to the numerical implementation of the fast TDBEM.

\subsubsection{Numerical treatment of the MMs, LCs, and the M2L operations}\label{s:fmm_remark_num}

Each of the seven MMs and LCs (as scalars) for a certain cell and time-interval can be treated as a $\Ps^3\Pt$-dimensional algebraic vector by concatenating all the components. For example, the scalar $M_{b,n}(\rmS,\rmJ)$ in (\ref{eq:dotphi_fmm}) can be represented as an algebraic vector $\mat{m}(\rmS,\rmJ)$ by defining
\begin{eqnarray*}
  \textrm{m}_{((b_1\Ps+b_2)\Ps+b_3)\Pt+n+1}(\rmS,\rmJ):=M_{b_1,b_2,b_3,n}(\rmS,\rmJ)
\end{eqnarray*}
for any $b_1$, $b_2$, $b_3\in[0,\Ps)$ and $n\in[0,\Pt)$. Similarly, the LC $L_{a,m}(\rmO,\rmI)$ can be expressed with $\mat{l}(\rmO,\rmI)$. Then, the M2L formula in (\ref{eq:dotphi_fmm}) can be written as a $\Ps^3\Pt$-dimensional matrix-vector product, i.e.
\begin{eqnarray}
  \mat{l}(\rmO,\rmI)=\mat{U}^d(\rmO,\rmS,\rmI,\rmJ)\mat{m}(\rmS,\rmJ),
  \label{eq:m2l_matrix}
\end{eqnarray}
where the M2L operator (matrix) $\mat{U}^d$ can be computed as
\begin{eqnarray*}
  \textrm{U}^d_{((a_1\Ps+a_2)\Ps+a_3)\Pt+m+1,((b_1\Ps+b_2)\Ps+b_3)\Pt+n+1}:=U^d_{a_1,a_2,a_3,b_1,b_2,b_3,m,n}(\rmO,\rmS,\rmI,\rmJ)
\end{eqnarray*}
for any $a_1$, $a_2$, $a_3$, $b_1$, $b_2$, $b_3\in[0,\Ps)$ and $m,n\in[0,\Pt)$.

The M2L operator $\mat{U}^d(\rmO,\rmS,\rmI,\rmJ)$ in (\ref{eq:m2l_matrix}) is precomputed for all the possible $\rmO$, $\rmS$, $\rmI$, and $\rmJ$. Because the function $U^d$ in (\ref{eq:U}) has a property of the transnational invariance with respect to space and time, the number of pairs of $\rmO$ and $\rmS$ is limited to a fixed number $316$ ($=7^3-3^3$). On the other hand, the number of pairs of $\rmI$ and $\rmJ$ increases as $\Nt$ increases. However, only $9$ pairs are enough by utilising the recurrence formulae, i.e. \cite[Formula~1]{takahashi2022}) in terms of the LC. Once $316\times 9$ M2L operators are computed and stored for a certain level, those for other levels can be obtained by a certain scaling because the cell size $\ds^{(l)}$ and the duration $\dt^{(l)}$ of level $l$ is halved (respectively, doubled) when $l$ increases (respectively, decreases) by one (recall Section~\ref{s:hierarchy}).

In addition, it should be noted that the matrix-vector product in (\ref{eq:m2l_matrix}) can be computed efficiently with the 4D fast Fourier transform (FFT), as mentioned in \cite[Section~4.5]{takahashi2014}.

\subsubsection{Parallelisation}\label{s:notes_parallel}

In the actual program, a multi-threaded parallelisation is considered with the help of OpenMP similarly to the acoustic program~\cite{takahashi2014}. Specifically, the loops over $\rmO$ at Line~\ref{line:near_O} of \refalgo{algo:near}, $\rmS$ at Line~\ref{line:upward_S} of \refalgo{algo:upward}, and $\rmO$ at Line~\ref{line:downward_O} of \refalgo{algo:downward} is annotated by the OpenMP directive `omp parallel for'. In addition, the same parallelisation is applied to the stage of precomputations, e.g. the computation of the $316\times 9$ M2L operators mentioned above.

\subsubsection{Program of the conventional TDBEM}

The pseudo program of the conventional TDBEM can be obtained by modifying that of the fast TDBEM. To this end, the loop over levels at Line~\ref{line:near_level} of \refalgo{algo:near} is limited to only the root level, i.e. $l=0$. Then, the cell $\rmO$ in the next line is the root cell only and its neighbour-list $\NL(\rmO)$ consists of $\rmO$ only. Therefore, the modified program can computes the interactions among all the $\Ne$ RWG bases. Then, the far-field computation at Line~\ref{line:main_far} of \refalgo{algo:main} is no longer necessary and thus removed.

\def\prtime#1{{\nprounddigits{0}\textrm{\numprint{#1}}}}
\def\prmem#1{{\nprounddigits{0}\npproductsign{\times}\textrm{\numprint{#1}}}}
\def\prnum#1{{\nprounddigits{2}\npproductsign{\times}\textrm{\numprint{#1}}}}

\section{Numerical examples}\label{s:num}

The proposed fast TDBEM was numerically checked through a test problem (Sections~\ref{s:test}--\ref{s:test_discuss}) and a demonstration (Section~\ref{s:demo}).

\subsection{Test problem}\label{s:test}

\begin{figure}[H]
  \centering
  \includegraphics[width=.35\textwidth]{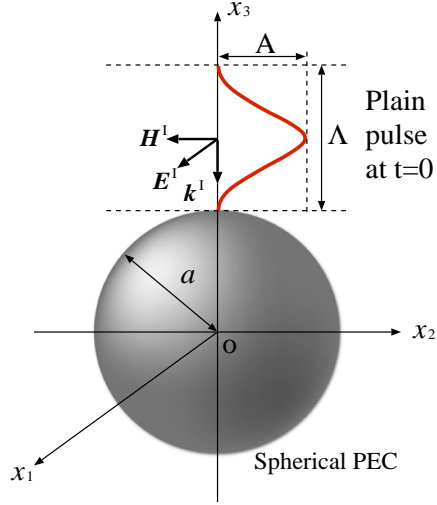}
  \caption{Configuration of the test problem.}
  \label{fig:test-config}
\end{figure}

Let us consider an EM scattering problem regarding a spherical PEC of radius $a=0.5$ at the origin in the free space of $\epsilon=\mu=1$ (Figure.~\ref{fig:test-config}). A plane incident wave that propagates in the $-x_3$-direction or $\bm{k}^{\mathrm{I}}:=(0,0,-1)^\mathrm{T}$ and has the following electric and magnetic fields was considered:
\begin{eqnarray*}
  \Einc(\bm{x},t)=\left(A\ \mathrm{Sin}^2\left(\frac{2\pi}{\Lambda}(ct+x_3-a)\right), 0, 0\right)^{\mathrm T},\quad
  \Hinc(\bm{x},t)=\frac{1}{\eta}\bm{k}^{\mathrm{I}}\times\Einc(\bm{x},t),
\end{eqnarray*}
where both the pulse length $\Lambda$ and the amplitude $A$ were given as $0.5$. Here, the function $\mathrm{Sin}()$ is defined as
\begin{eqnarray*} 
  \mathrm{Sin}(x):=\begin{cases}
  \sin(x) & 0\le x\le 2\pi\\
  0 & \textrm{otherwise}
  \end{cases}.
\end{eqnarray*}

The proposed fast TDBEM was compared with the conventional one. In both methods, the coupling parameters $\theta$ and $\zeta$ of the CFIE in (\ref{eq:cfie}) were simply selected as one.

\begin{figure}[H]
  \centering
  \includegraphics[width=.4\textwidth]{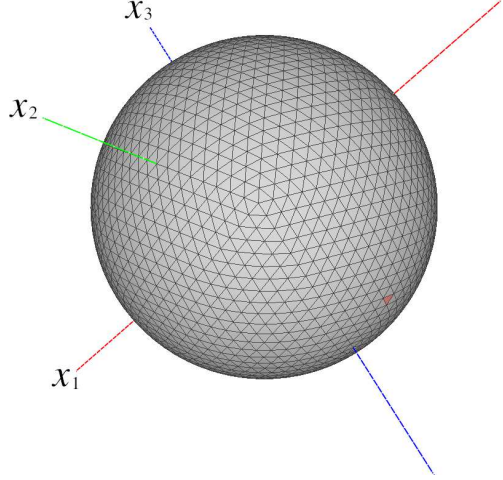}
  \caption{Mesh for Case 3 ($5120$ triangles; $7680$ edges). The red triangle is the element whose profile of the current density $\bm{J}$ is sampled in Figure~\ref{fig:test-profile}.}
  \label{fig:h03-e4300}
\end{figure}

Seven cases of mesh or $\Ns$ ($\equiv 2\Ne/3$) were considered as in Table~\ref{tab:test}; Figure~\ref{fig:h03-e4300} shows the mesh for Case 3. Since the size of the PEC is fixed, which is consistent to Assumption~\ref{assume:S}, edge lengths, denoted by $\Ds$, decrease as $\Ns$ increases, as shown in the table. Further, the maximum level $\maxlev$ of the octree varied from $2$ to $5$ by setting $\Nleaf=100$. For all these cases, the time-step size $\Dt$ was fixed in order to be consistent to Assumption~\ref{assume:Dt} and selected as $0.005$. In this case, the breakdown of the IFMM, mentioned in Remark~\ref{remark:breakdown}, never occurred. Actually, the number $\Mt^{(\maxlev)}$ of time-steps per time-interval at level $\maxlev$ was determined as in the table.

The present distribution of the edges (RWG bases) is close to the plane case in (\ref{eq:assume4}) rather than the uniform one in (\ref{eq:assume8}) of Assumption~\ref{assume:48}. For an example of Case 5, where $\maxlev$ is $5$ when $\Nleaf=100$, the LHSs of (\ref{eq:assume8}) and (\ref{eq:assume4}) are computed as $\Nleaf\cdot 8^\maxlev=100\cdot 8^5=3276800$ and  $\Nleaf\cdot 4^\maxlev=100\cdot 4^5=102400$, respectively. Clearly, the latter is closer to the actual number of edges, i.e. $\Ne\equiv 3\Ns/2=122880$, than the former.

\begin{table}[H]
  \begin{center}
    \caption{Seven cases for the test problem.}
    \label{tab:test}
    \begin{tabular}{|c|r|c|c|c|c|c|}
      \hline
      Case & $\Ns$ & \multicolumn{3}{c|}{$\Ds$} & $\maxlev$ & $\Mt^{(\maxlev)}$\\
      &  & Min. & Ave. & Max. & &\\
      \hline
      1 &  1280 & 0.069 & 0.075 & 0.084 & 2 & 50\\
      2 &  2880 & 0.046 & 0.050 & 0.056 & 2 & 50\\
      3 &  5120 & 0.035 & 0.038 & 0.042 & 3 & 25\\
      4 & 11520 & 0.023 & 0.025 & 0.028 & 4 & 12\\
      5 & 20480 & 0.017 & 0.019 & 0.021 & 4 & 12\\
      6 & 46080 & 0.012 & 0.013 & 0.014 & 5 &  6\\
      7 & 81920 & 0.009 & 0.010 & 0.011 & 5 &  6\\
      \hline
    \end{tabular}
  \end{center}
\end{table}

To see the trend of the accuracy and runtime of the fast TDBEM, three cases of $\Ps$ and $\Pt$, which denote the number of interpolation points (recall Section~\ref{s:fmm_expand_U}), were tested. Specifically, let both $\Ps$ and $\Pt$ take the value of $4$, $6$, or $8$. According to these values, the fast TDBEM is referred to as FAST4, FAST6, and FAST8, while the conventional TDBEM is referred to as CONV.

For the present problem, it is possible to calculate a reference solution, say $\bm{J}^{\mathrm{reference}}$, semi-analytically. To this end, one first calculates the exact solution in the Laplace transformed domain with the help of the solution in the frequency or Fourier-transformed domain~\cite{bowman1987} and then applies a numerical inverse Laplace transform to the exact solution.

To measure the accuracy of the both TDBEMs, their errors relative to $\bm{J}^{\mathrm{reference}}$ were calculated with the following relative $l^2$-error:
\begin{eqnarray}
  \textrm{Relative $l^2$-error}:=\frac{\displaystyle\sum_{i=1}^{\Ns}\sum_{\alpha=1}^{\Nt-1}\left|\bm{J}^{\mathrm{TDBEM}}(\overline{S}_i,t_\alpha)-\bm{J}^{\mathrm{reference}}(\overline{S}_i,t_\alpha)\right|^2}{\displaystyle\sum_{i=1}^{\Ns}\sum_{\alpha=1}^{\Nt-1}\left|\bm{J}^{\mathrm{TDBEM}}(\overline{S}_i,t_\alpha)\right|^2},
  \label{eq:error}
\end{eqnarray}
where $\overline{S}_i$ denotes the centre of triangle $S_i$ and $\bm{J}^{\mathrm{TDBEM}}$ stands for the current density obtained from (\ref{eq:J}) by either the conventional or fast TDBEM.

In all the computations, a workstation with Intel's Xeon CPU (model: Gold 6250, clock rate: 3.90~GHz, number of computing cores: 16) and 768~GB memory was used. The both TDBEM programs are parallelised by using the OpenMP as mentioned in Section~\ref{s:notes_parallel}.

\subsection{Results}\label{s:test_results}

The proposed fast TDBEM was assessed with regard to the accuracy, computation time, and memory consumption. It should be noted that the conventional method could not solve Cases 4--7 because it ran out of memory on the way of the computations.

\begin{figure}[H]
  \centering
  \includegraphics[width=.55\textwidth]{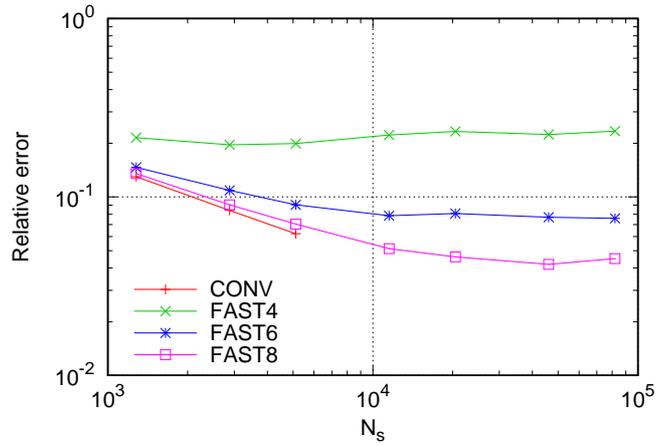}
  \caption{Relative $l^2$-error for the test problem.}
  \label{fig:test-error}
\end{figure}

Figure~\ref{fig:test-error} compares the relative error of the both TDBEMs. The error of the conventional method decreased monotonically as $\Ns$ increased. This can be interpreted as the decrease of the discretisation error. On the other hand, for each precision level, the error of the fast method decreased and then saturated with $\Ns$. This behaviour is probably due to the approximation error associated with the choice of $\Ps$ and $\Pt$. Further, the accuracy tends to improve as $\Ps$ and $\Pt$ increase for every $\Ns$.

\begin{figure}[H]
  \centering
  \includegraphics[width=.55\textwidth]{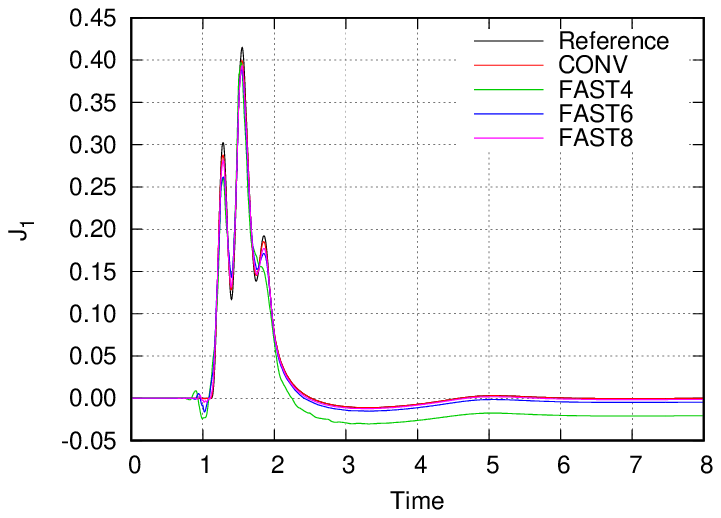}\\
  \includegraphics[width=.55\textwidth]{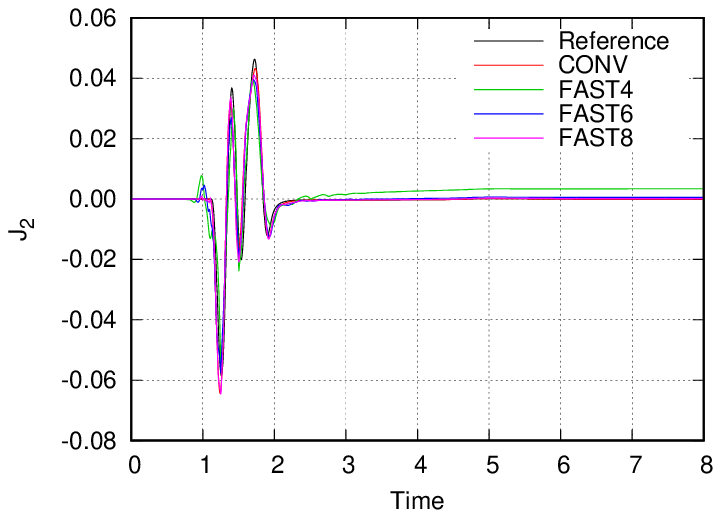}\\
  \includegraphics[width=.55\textwidth]{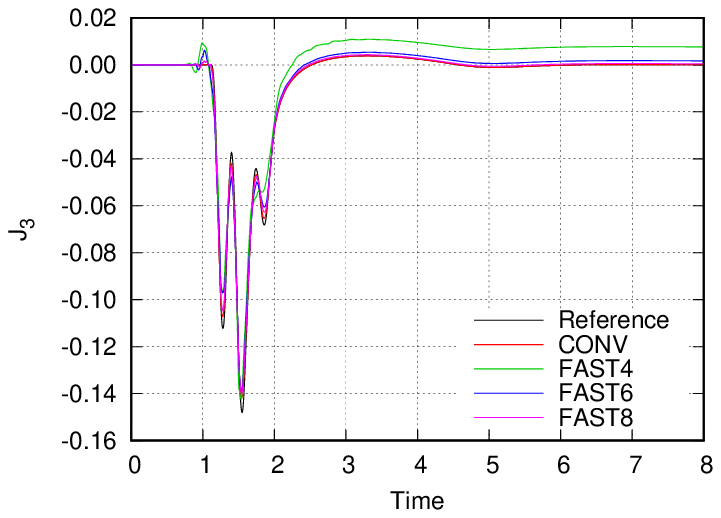}
  \caption{Profile of $\bm{J}=(J_1,J_2,J_3)^{\mathrm{T}}$ at the centre of the triangle coloured in red in Figure~\ref{fig:h03-e4300}.}
  \label{fig:test-profile}
\end{figure}

The relative errors were around $10^{-1}$ or $10\%$ in Figure~\ref{fig:test-error}. To check how the corresponding current densities $\bm{J}$ look, Figure~\ref{fig:test-profile} plots the three components of $\bm{J}$ at the centre of a randomly-selected element, which is coloured in red in Figure~\ref{fig:h03-e4300}, for Case 3. As observed, FAST4 is obviously inaccurate, while FAST6 and FAST8 are better except the parts where the reference solution changes sharply. It should be noted that the relative $l^2$-errors of the underlying $\bm{J}$ were \prnum{5.698574e-02}, \prnum{2.711700e-01}, \prnum{1.079286e-01}, and \prnum{7.232439e-02} for CONV, FAST4, FAST6, and FAST8, respectively. These values are close to those in Figure~\ref{fig:test-error} at $\Ns=5120$. So, the presented profiles can be considered as the representatives of all the profiles.

\begin{figure}[H]
  \centering
  \includegraphics[width=.55\textwidth]{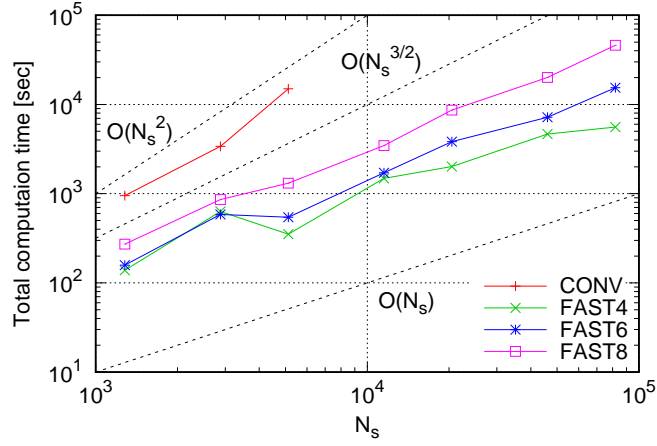}
  \caption{Total computation time for the test problem.}
  \label{fig:test-total}
\end{figure}

Figure~\ref{fig:test-total} shows the total computation time. First of all, the proposed method outperformed the conventional one in Cases 1, 2, and 3, which would be true for the larger-size cases. From Figures~\ref{fig:test-total} and \ref{fig:test-error}, the trade-off between computation time and error is observed for the fast method. Further, Figure~\ref{fig:test-total} clearly shows that the present IFMM can reduce the computational complexity from $\Order(\Ns^2)$ to $\Order(\Ns^{3/2})$, which is for the plane-like distribution of edges in space.

\begin{figure}[H]
  \centering
  \includegraphics[width=.55\textwidth]{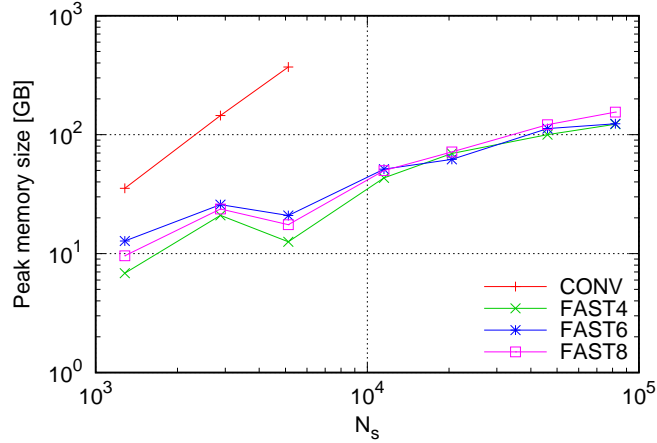}
  \caption{Memory consumption for the test problem.}
  \label{fig:test-memory}
\end{figure}

Figure~\ref{fig:test-memory} plots the memory consumption of the both TDBEMs. For brevity, the peak memory size was measured.\footnote{On a Linux system, the peak memory size or `peak resident set size' of a process can be known by the entry `\texttt{VmHWM}' in the status file for the corresponding UNIX process.} The conventional method stores the non-zero entries of the coefficient matrices $\mat{Z}^{(1)}$, $\mat{Z}^{(2)}$, and so on. On the other hand, the fast method stores a part of the non-zero entries regarding the near-field computation (i.e. $\mat{Z}_{\rm near}^{(\gamma)}$), as explained in Section~\ref{s:pseudo}. Instead, the far-field computation needs to store some quantities such as the MMs and LCs for every cell as well as the M2L operators. However, for reasonable values of $\Ps$ and $\Pt$, the far-field computation uses less memory than computing $\mat{Z}_{\rm far}^{(\gamma)}$ ($=\mat{Z}^{(\gamma)}-\mat{Z}_{\rm near}^{(\gamma)}$) directly. As a result, the IFMM can save the memory significantly.

Overall, the numerical results presented here are deemed satisfactory to validate the usefulness of the proposed fast TDBEM.

\subsection{Discussions}\label{s:test_discuss}

\subsubsection{Comparison with the conventional CFIEs}

The present TDBEM indeed requires both the MFIE and $\DtMFIE$. To see this, the following three types of CFIE were compared for both the conventional and fast algorithms:
\begin{itemize}
\item $\text{CFIE1} = \partial_t\text{EFIE} + \partial_t\text{MFIE} + \text{MFIE}$; this is exactly the CFIE in (\ref{eq:cfie}),
  
\item $\text{CFIE2}=\partial_t \text{EFIE} + \partial_t\text{MFIE}$,
  
\item $\text{CFIE3}=\partial_t \text{EFIE} + \text{MFIE}$.

\end{itemize}

\begin{figure}[H]
  \centering
  \includegraphics[width=.55\textwidth]{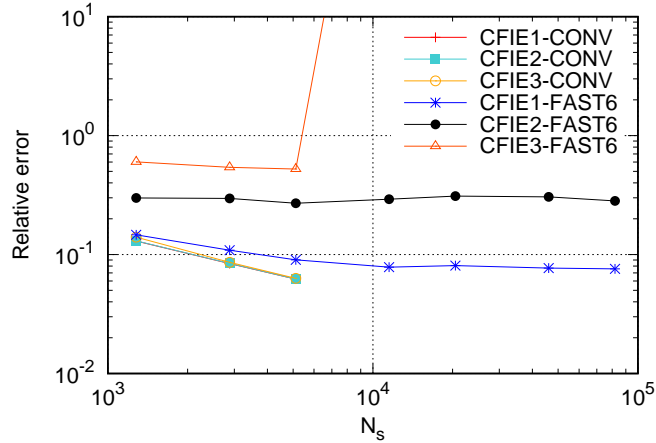}
  \caption{Relative error for the three types of CFIE. The results of CFIE1-CONV and CFIE1-FAST6 are the same as those of CONV and FAST6 in Figure~\ref{fig:test-error}, respectively. The error of CFIE3-FAST6 was significantly large at larger $\Ns$s and thus not plotted.}
  \label{fig:test-error-discuss}
\end{figure}

Figure~\ref{fig:test-error-discuss} compares the six cases of the TDBEM in terms of the relative $l^2$-error in (\ref{eq:error}). In the case of the conventional algorithm, CFIE2 was almost the same as CFIE1 and CFIE3 was slightly worse than the others. This tendency was basically true also for the fast algorithm, but its intrinsic approximation (owing to the separation of variables by interpolation) seems to amplify the error of the corresponding conventional methods. In particular, CFIE3 caused the late time instability. It is difficult to justify the necessity of CFIE1 mathematically, but this comparison indicates that using the MFIE and $\partial_t\text{MFIE}$ simultaneously is optional for the conventional TDBEM but indispensable for the fast TDBEM.

Moreover, the choices of the coupling parameters $\theta$ and $\zeta$ in the CFIE (\ref{eq:cfie}) can influence the error result. It would be possible to seek the optimal values, but this study did not pursue them.

\subsubsection{Choice for temporal differentiation}

It would be interesting to check other possibilities of the second-order temporal-differentiation $\partial_t^2$ acting on the vector potential $\bm{A}$ in the $\DtEFIE$. In the present case of (\ref{eq:ddotA_fmm}), both the kernel $U^d$ and the interpolant $\Lag_m$ are differentiated once. In this case, the interpolation of the derivative $\dot{U}^d\left(\equiv cdU^{d-1}\right)$ is also necessary in addition to that of $U^d$. This requires an additional computation time and memory at the stage of the precomputing, but they are not significant. On the other hand, it is also possible to apply $\partial_t^2$ either $U^d$ or $\Lag_m$. These two cases were tested with maintaining $d=2$ (i.e. the quadratic B-spline temporal basis) and the interpolant (i.e. the CHI using a finite difference approximation) as well as $\Ps=\Pt=6$ (i.e. FAST6).

\begin{figure}[H]
  \centering
  \includegraphics[width=.55\textwidth]{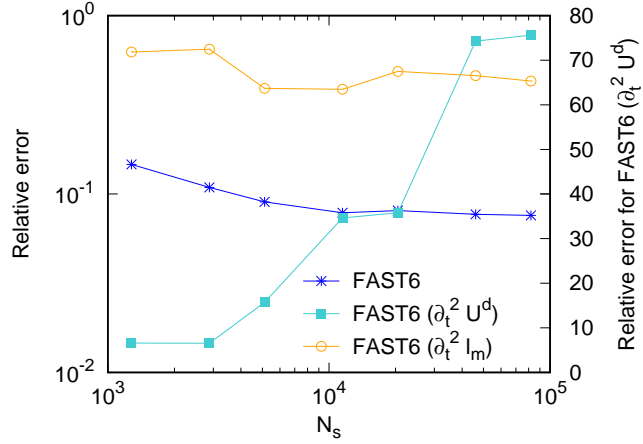}
  \caption{Relative error for the three patterns of the temporal differentiations regarding the vector potential $\bm{A}$. The vertical axis in the RHS is only for `FAST6 ($\partial_t^2 U^d$)' and shows the exponent of the relative error in the logarithmic scale; e.g., '20' means the relative error of $10^{20}$. The result of FAST6 is the same as that in Figure~\ref{fig:test-error}.}
  \label{fig:test-error-discuss2}
\end{figure}

The relative $l^2$-error is shown in Figure~\ref{fig:test-error-discuss2}, where `FAST6 ($\partial_t^2U^d$)' and `FAST6 ($\partial_t^2\Lag_m$)' are results for using $\partial_t^2U^d$ and $\partial_t^2\Lag_m$, respectively. In the case of $\partial_t^2U^d$, FAST6 resulted in a late time-instability in every $\Ns$ and thus the corresponding error became significantly large. 
This is probably because the derivative $\partial^2 U^d$ is discontinuous owing to $d=2$ and, therefore, the interpolated derivative is erroneous. The accuracy could be improved by using a larger $d\ge 3$. However, using $d\ge 3$ makes the algorithm unstable; even the conventional TDBEM is unstable with $d\ge 3$.

The result of FAST6 using $\partial_t^2\Lag_m$ was better than that of $\partial_t^2 U^d$ but worse than the original FAST6, based on $\partial_tU^d$ and $\partial_t\Lag_m$, as observed in the same figure. In this case, the differentiated interpolant $\partial_t^2\Lag_m$ is piece-wise linear because the $\Lag_m$ is a cubic polynomial. However, the smoothness would be insufficient to express the true solution that behaves smoothly and sharply as seen in Figure~\ref{fig:test-profile}.

\subsection{Demonstration}\label{s:demo}

\begin{figure}[hbt]
  \centering
  \includegraphics[width=.7\textwidth]{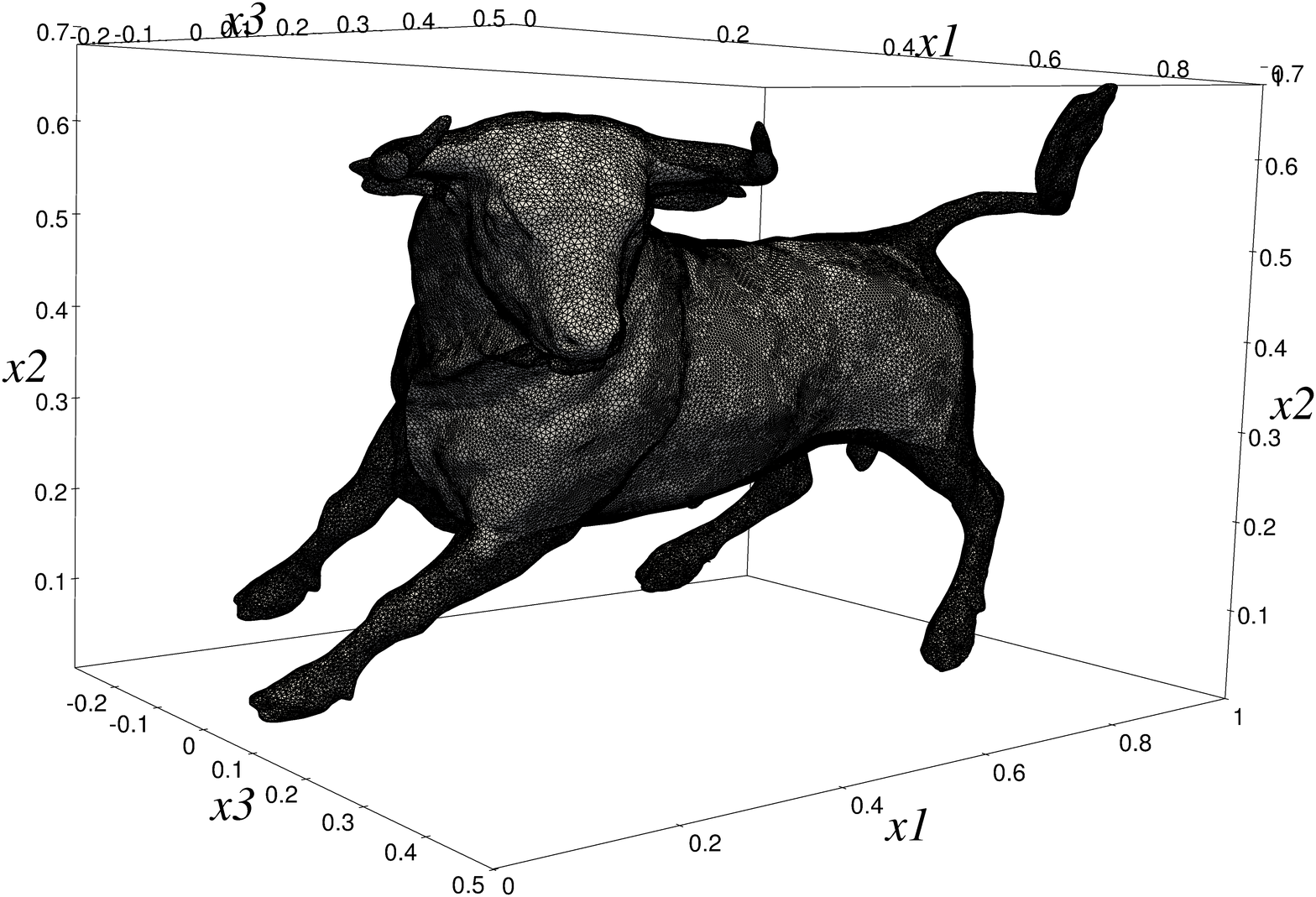}\\[\baselineskip]
  \includegraphics[width=.4\textwidth]{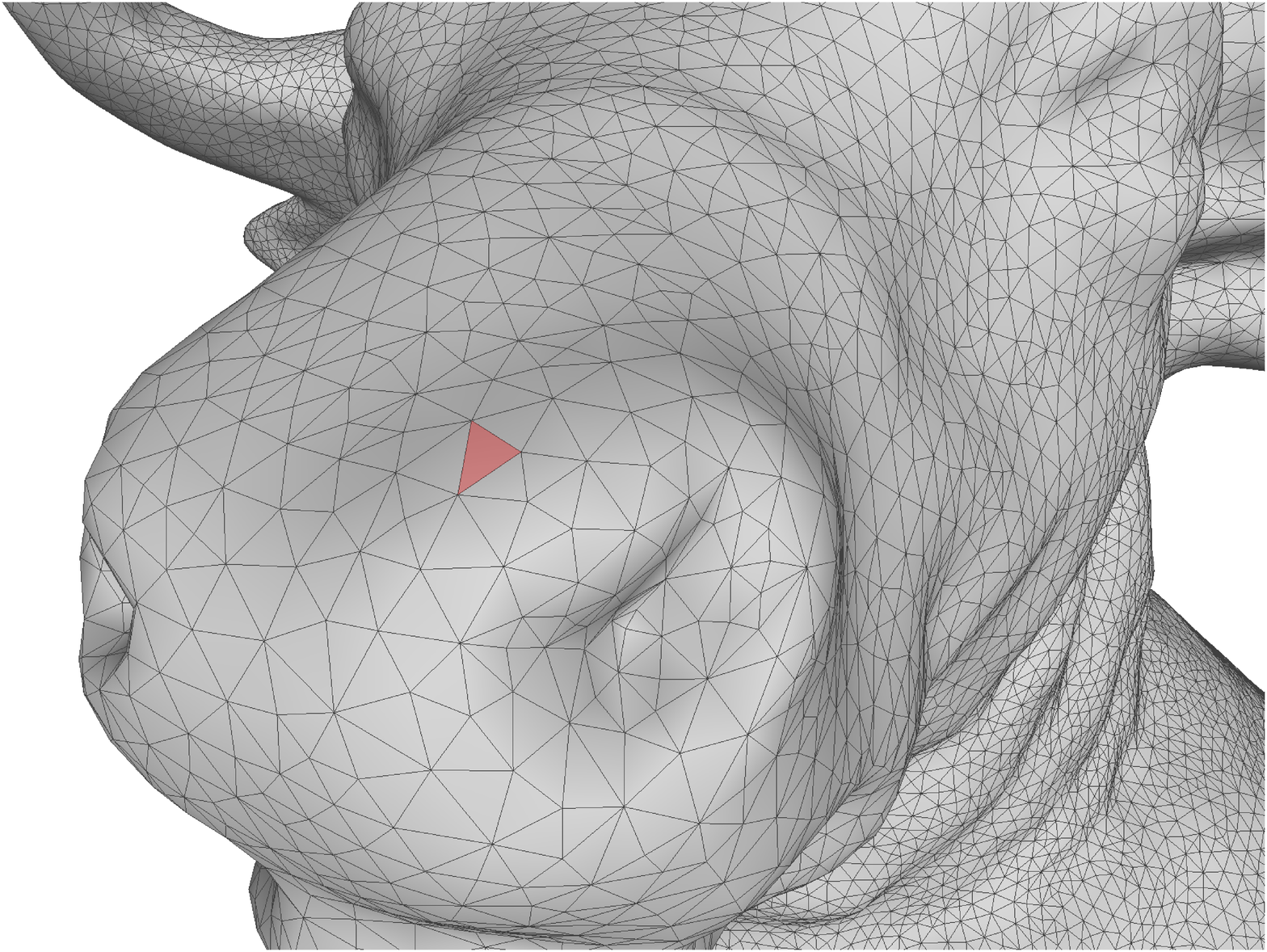}
  \caption{Mesh for the `bull' model consisting of 139282 triangles, that is, 208923 edges: (Top) Entire of the model. The incident EM wave propagates in $-x_3$-direction. (Bottom) Head of the model. At the centre of the red triangle on the nose, the profile of current density $\bm{J}$ is calculated; see Figure~\ref{fig:bull03-profile}.}
  \label{fig:bull03}
\end{figure}

To see the applicability of the proposed fast TDBEM, a more complex and large-scale `bull' model, which is placed in $[0.0,1.0]\otimes[0.0,0.7]\otimes[-0.2,0.5]$, was considered. Figure~\ref{fig:bull03} shows the mesh consisting of $139282$ ($=\Ns$) boundary elements: the minimum and maximum length of the edges are \prnum{6.980593e-04} and \prnum{7.548610e-03}, respectively.\footnote{The original mesh (`\texttt{bull.off}') was obtained from the source code of the Computational Geometry Algorithm Library (CGAL) (\url{https://www.cgal.org/index.html}) and then refined with the mesh-processing software MeshLab (\url{https://www.meshlab.net/}).}  Also, $\Dt=0.0025$, $\Nt=1600$, and FAST6 were selected. In addition, the incident wave was given as a Gaussian plane pulse that propagates in $-x_3$-direction, i.e. $\bm{k}^{\mathrm{I}}=(0,0,-1)^\mathrm{T}$, and characterised with
\begin{eqnarray*}
  \Einc(\bm{x},t)=\left(\frac{1}{\sqrt{2\pi\sigma^2}}\exp\left(-\frac{(t+(x_3-0.5)/c-6\sigma)^2}{2\sigma^2}\right),0,0\right)^{\mathrm{T}},\quad
  \Hinc(\bm{x},t)=\frac{1}{\eta}\bm{k}^{\mathrm{I}}\times\Einc(\bm{x},t),
\end{eqnarray*}
where $\sigma=0.05$ and $\epsilon=\mu=1$ were assumed.

\begin{figure}[hbt]
  \centering
  \iffalse 
  \begin{tabular}{ccc}
    \includegraphics[width=.3\textwidth]{figure/220905/demo2/X0120_fix.eps}&
    \includegraphics[width=.3\textwidth]{figure/220905/demo2/X0150_fix.eps}&
    \includegraphics[width=.3\textwidth]{figure/220905/demo2/X0180_fix.eps}\\
    $t=120\Dt$ & $t=150\Dt$ & $t=180\Dt$\\
    \includegraphics[width=.3\textwidth]{figure/220905/demo2/X0210_fix.eps}&
    \includegraphics[width=.3\textwidth]{figure/220905/demo2/X0240_fix.eps}&
    \includegraphics[width=.3\textwidth]{figure/220905/demo2/X0270_fix.eps}\\
    $t=210\Dt$ & $t=240\Dt$ & $t=270\Dt$\\
    \includegraphics[width=.3\textwidth]{figure/220905/demo2/X0300_fix.eps}&
    \includegraphics[width=.3\textwidth]{figure/220905/demo2/X0330_fix.eps}&
    \includegraphics[width=.3\textwidth]{figure/220905/demo2/X0360_fix.eps}\\
    $t=300\Dt$ & $t=330\Dt$ & $t=360\Dt$\\
  \end{tabular}
  \else 
  \begin{tabular}{ccc}
    \includegraphics[width=.3\textwidth]{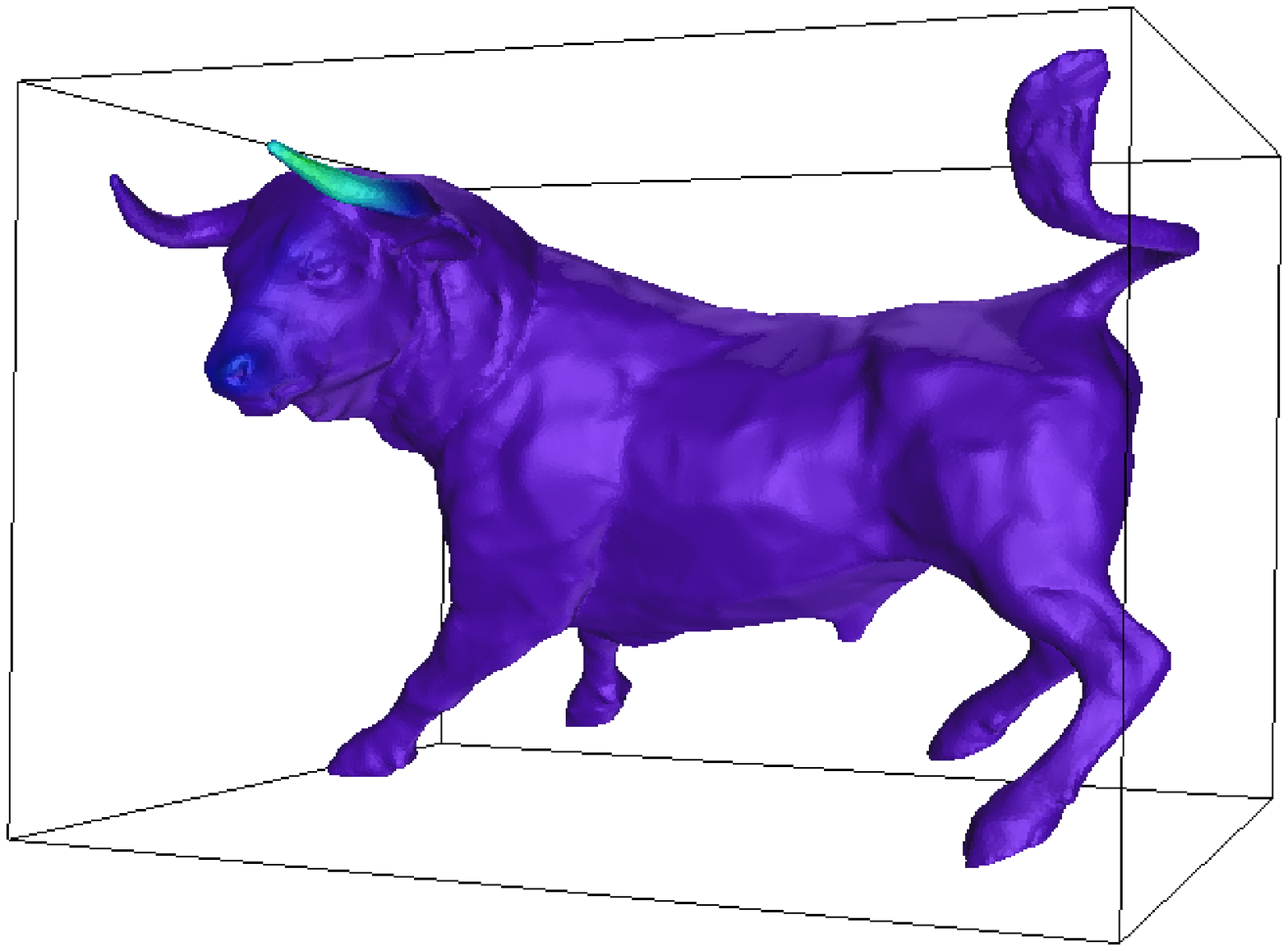}&
    \includegraphics[width=.3\textwidth]{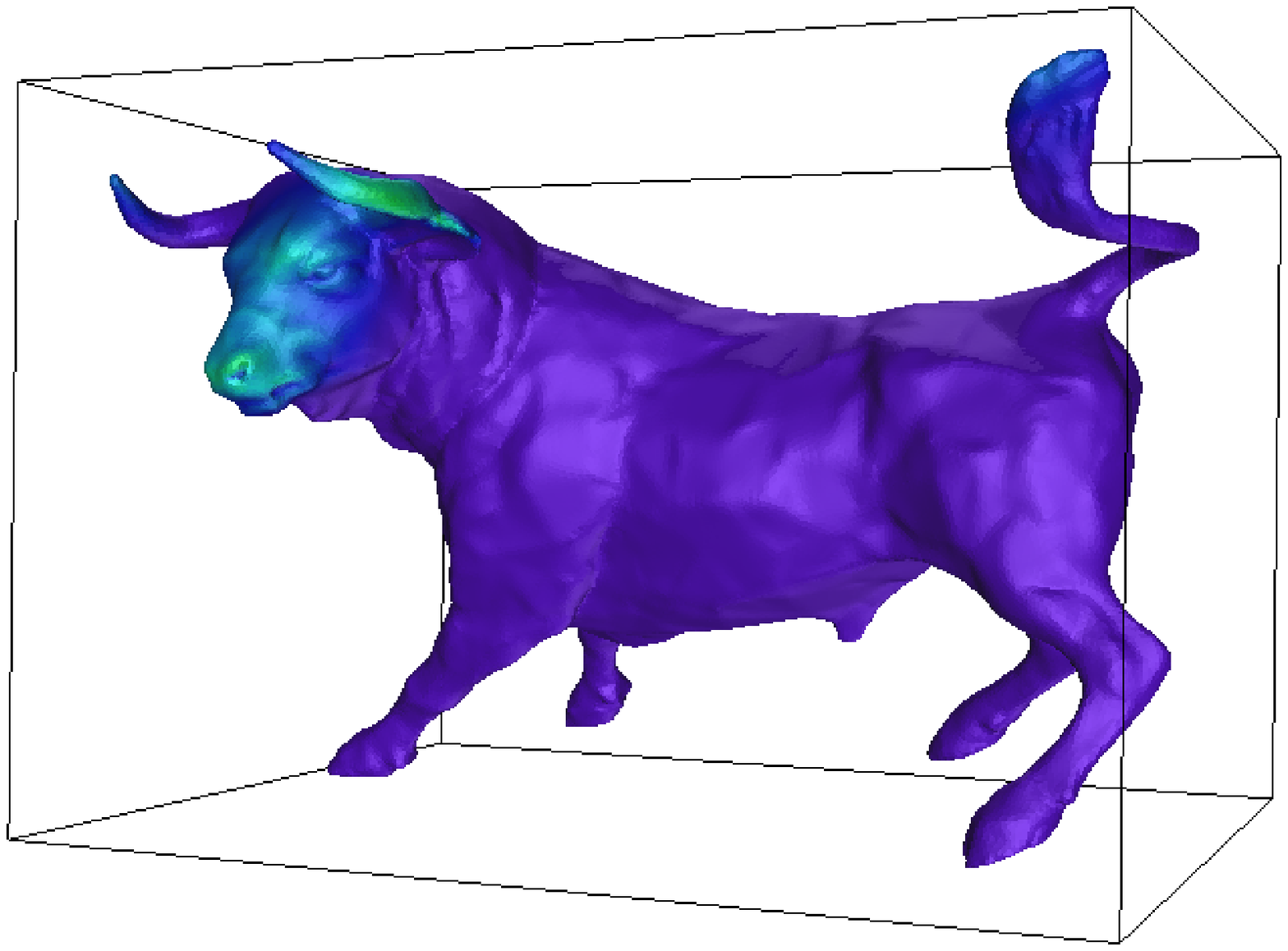}&
    \includegraphics[width=.3\textwidth]{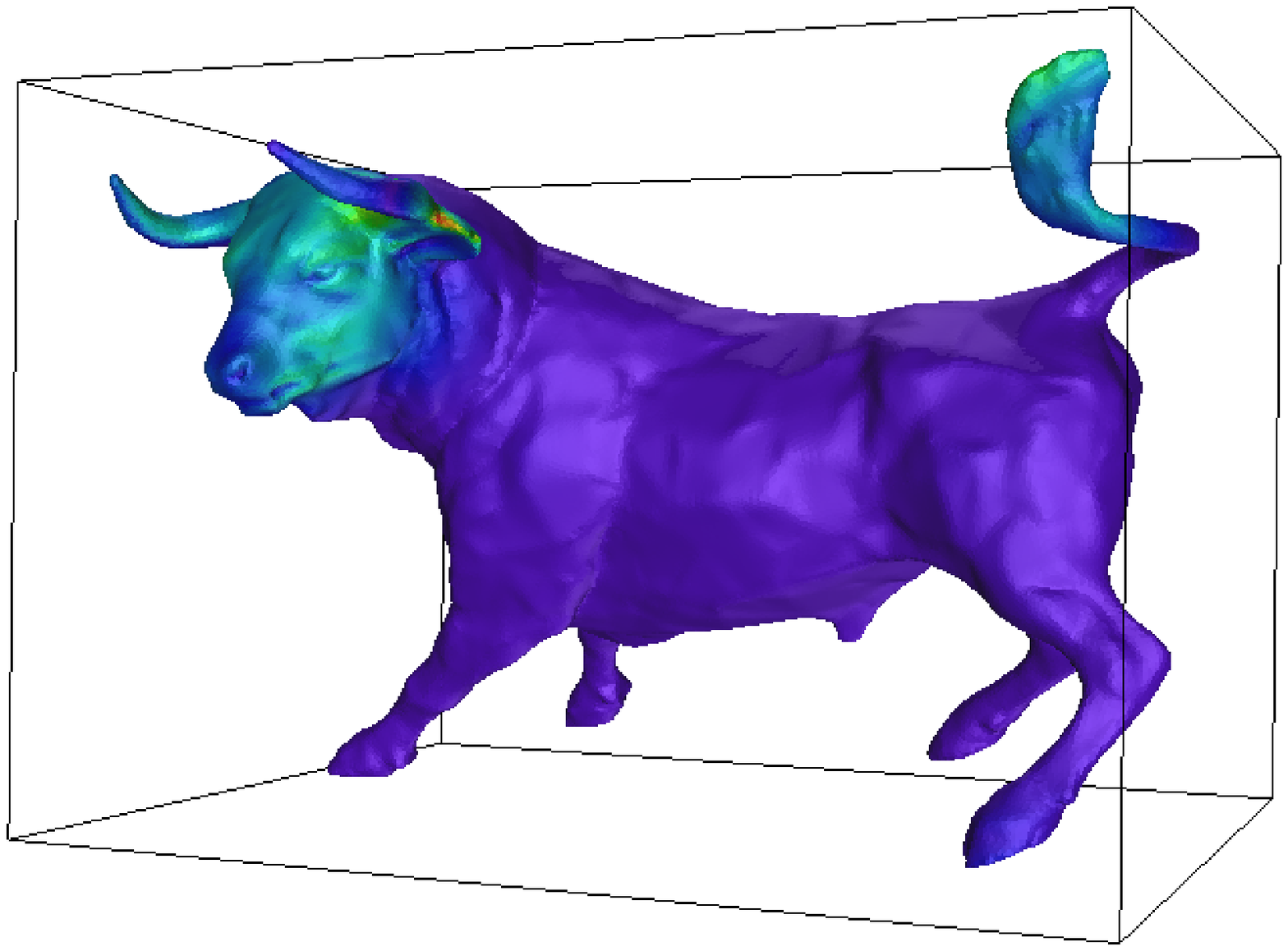}\\
    $t=0.300$ & $t=0.375$ & $t=0.450$\\
    \includegraphics[width=.3\textwidth]{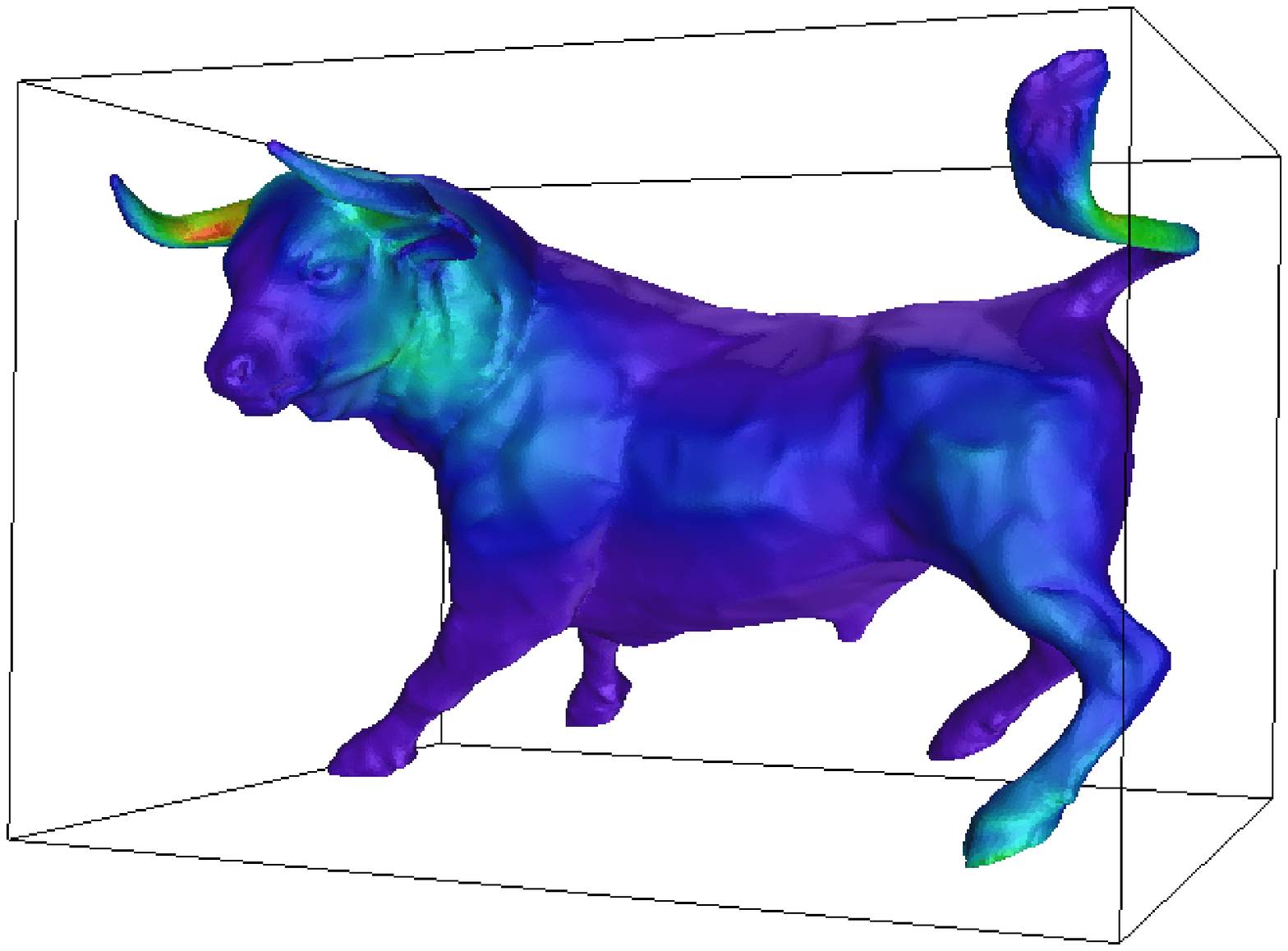}&
    \includegraphics[width=.3\textwidth]{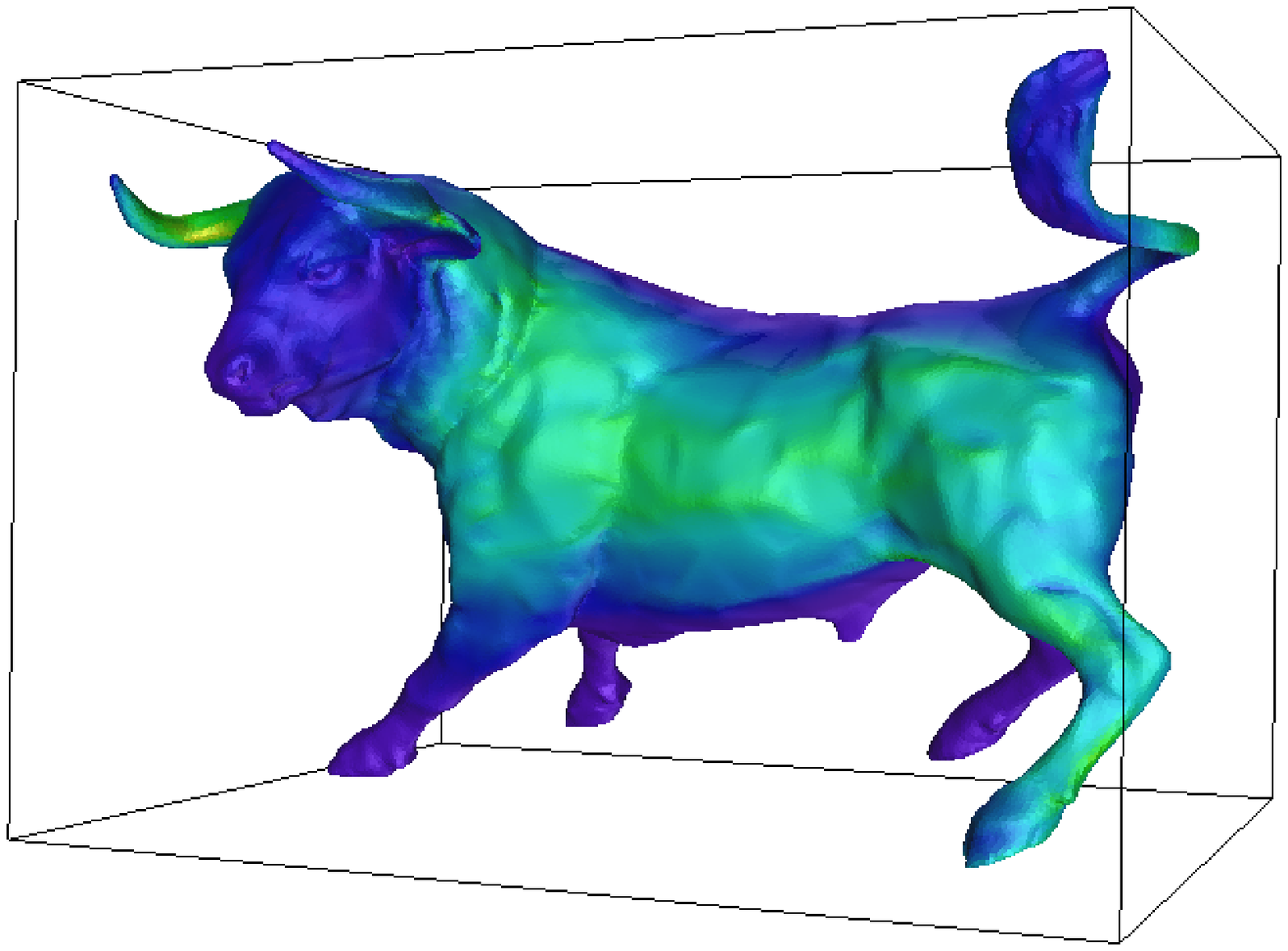}&
    \includegraphics[width=.3\textwidth]{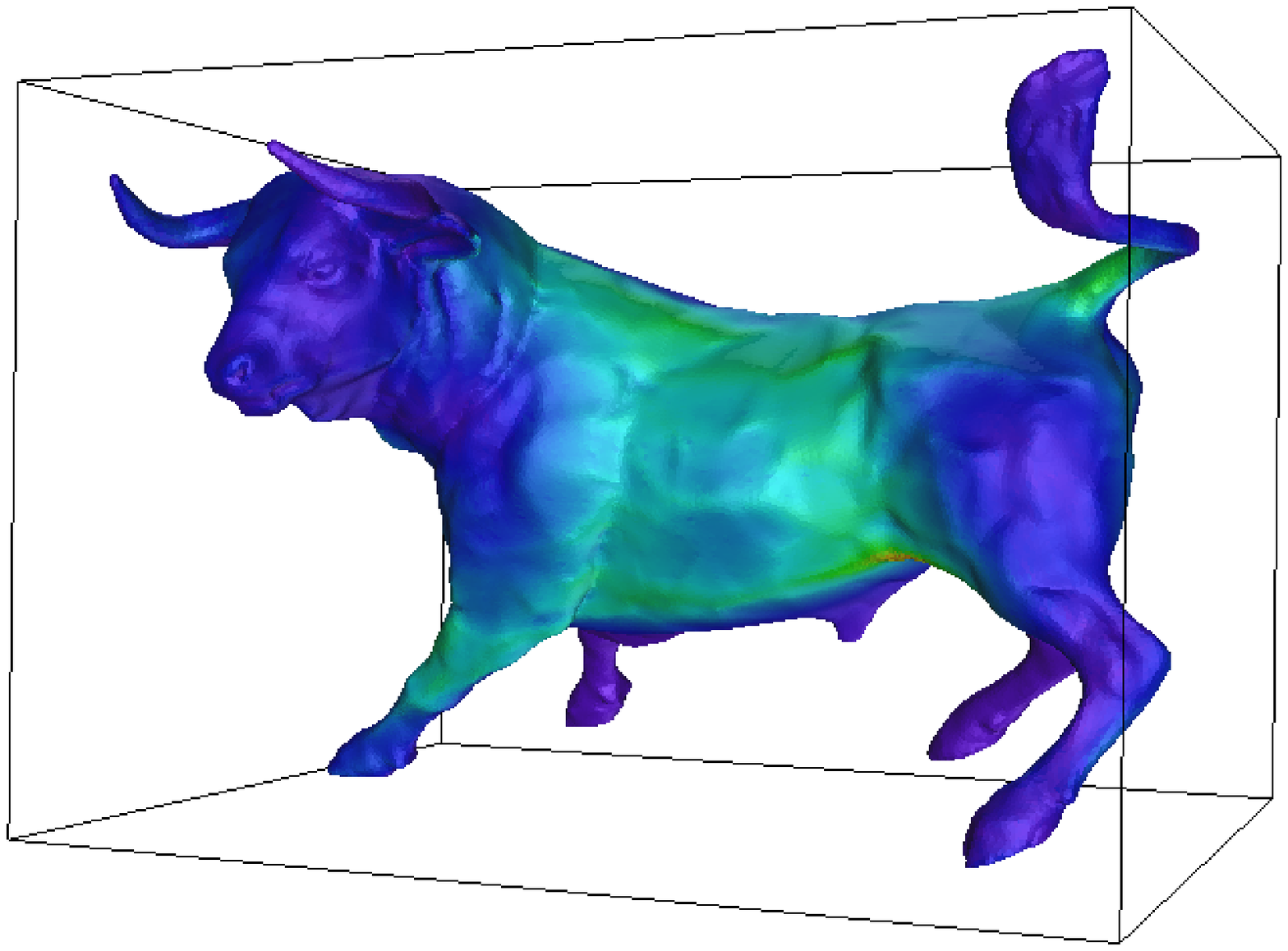}\\
    $t=0.525$ & $t=0.600$ & $t=0.675$\\
    \includegraphics[width=.3\textwidth]{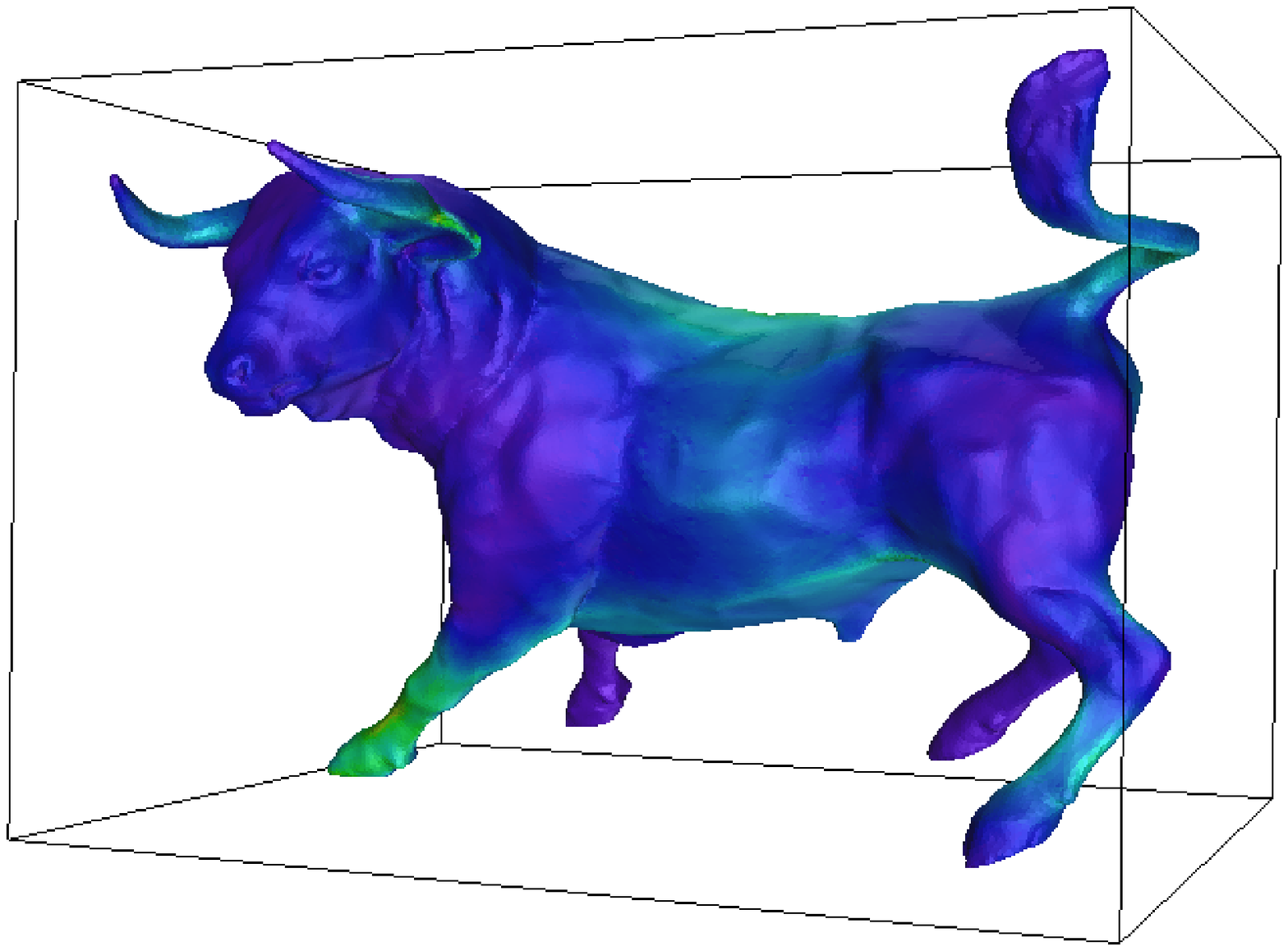}&
    \includegraphics[width=.3\textwidth]{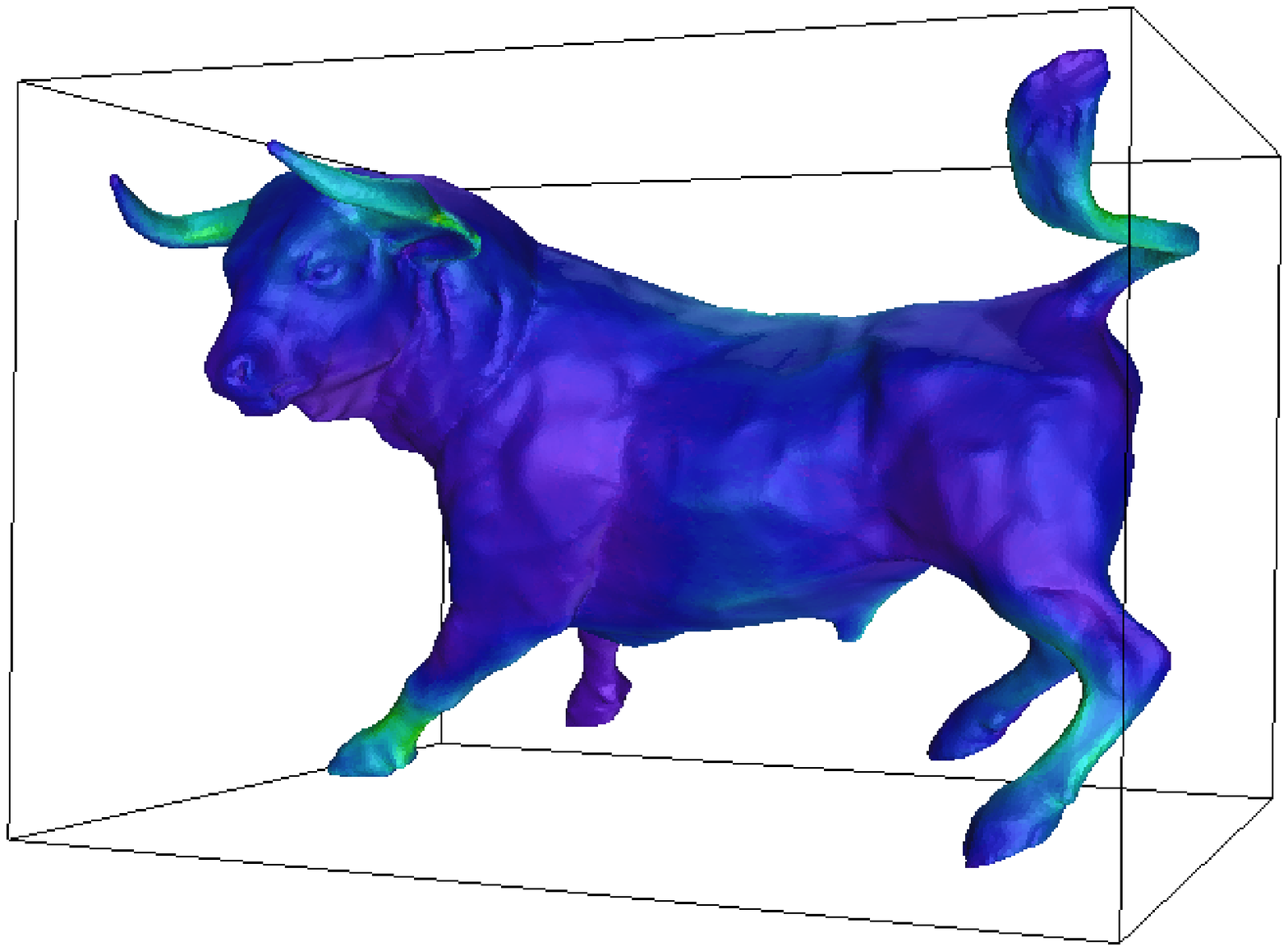}&
    \includegraphics[width=.3\textwidth]{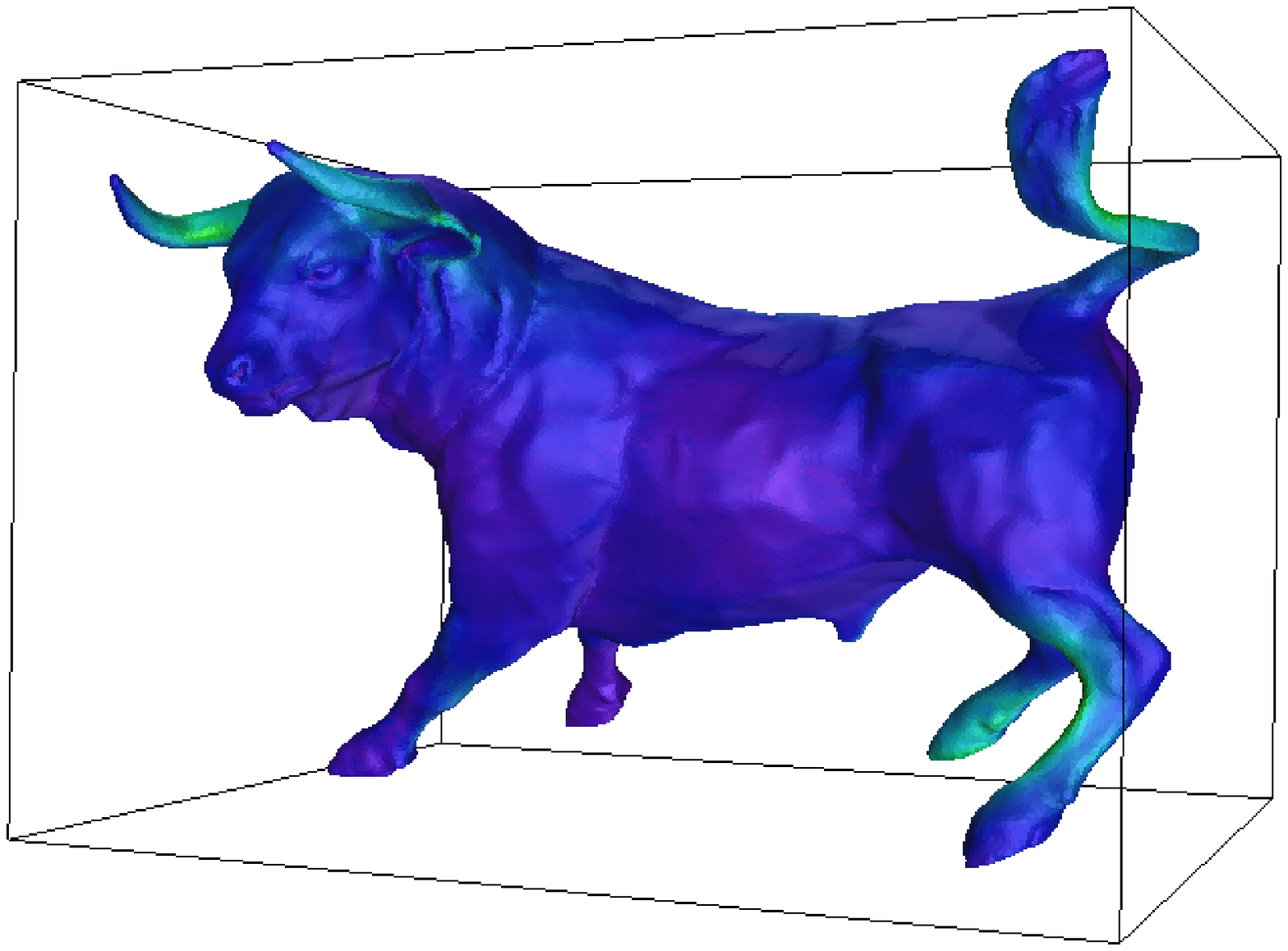}\\
    $t=0.750$ & $t=0.825$ & $t=0.900$\\
  \end{tabular}
  \fi
  \includegraphics[width=.3\textwidth]{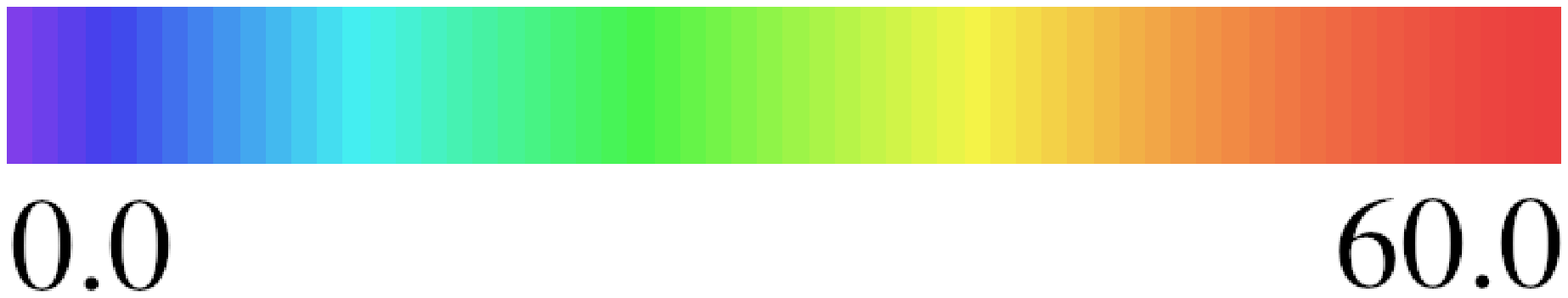} 
  \caption{Distribution of $|\bm{J}|$ on the model for selected time-steps. The incident EM wave propagates from the front side to the back side of the box.}
  \label{fig:bull03-J}
\end{figure}

\begin{figure}[hbt]
  \centering
  \includegraphics[width=.7\textwidth]{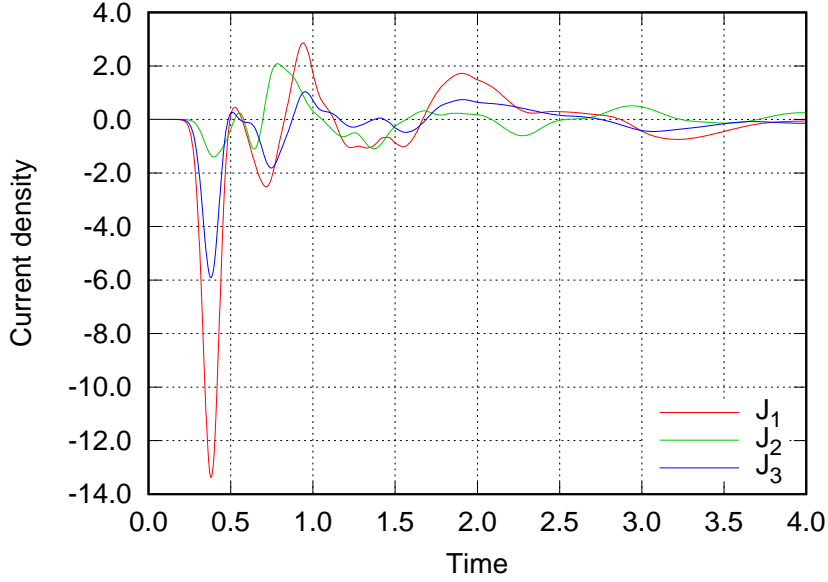}
  \caption{Profile of $\bm{J}=(J_1,J_2,J_3)^{\mathrm{T}}$ at the centre of the red triangle in Figure~\ref{fig:bull03}(bottom).}
  \label{fig:bull03-profile}
\end{figure}

Figure~\ref{fig:bull03-J} shows the magnitude of the surface current density $\bm{J}$ for selected time-steps. Moreover, Figure~\ref{fig:bull03-profile} shows the profile of $\bm{J}$ at the centre of the red triangle in Figure~\ref{fig:bull03}(bottom). From Figure~\ref{fig:bull03-profile}, the peak of $\bm{J}$ is around $t=0.5$, which is consistent to the result in Figure~\ref{fig:bull03-J}.

\def\prnum#1{{\nprounddigits{1}\npproductsign{\times}\textrm{\numprint{#1}}}}

The total computation time was \prnum{9.93375159777777777777} hour and the memory consumption was \prnum{277.331916} GB with the same workstation used in Section~\ref{s:test}.

\section{Conclusion}\label{s:conclusion}

The present study proposed an IFMM to accelerate the TDBEM for EM scattering problems in 3D regarding PECs. The targeted TDBEM is based on an unconventional CFIE as it considers both the MFIE and its temporal differentiation in addition to the time-differentiated EFIE. The accuracy of the CFIE is numerically rationalised as long as the surface current density is discretised with the RWG basis for space and the quadratic B-spline basis for time. Formulating the IFMM for the CFIE can be done by following the previous studies on acoustics~\cite{takahashi2014,takahashi2022}. The major difference from the acoustic case is in the number of MMs and LCs, some of which are vectors rather than scalars. Hence, they can increase the amount of computations to some extent. Nevertheless, the overall algorithm for the EM case is essentially the same as that for the acoustic case. Accordingly, the IFMM-accelerated TDBEM for electromagnetics possesses the complexity of $\Order(\Ns^{1+\delta}\Nt)$, where $\delta$ is typically estimated as $1/2$ or $1/3$ under Assumptions~\ref{assume:S}--\ref{assume:48}.

The numerical analyses in Section~\ref{s:num} investigated the proposed fast TDBEM with regard to its performance, accuracy, memory usage, and feasibility to a large-scale problem such as $\Ns=139282$ and $\Nt=1600$. The results are satisfactory and some additional computations discussed the validity for the formulation of the proposed method.

The future plans include a theoretical enhancement from PECs to penetrable bodies, an accuracy improvement by exploring a more appropriate interpolant beyond the CHI using finite-difference-approximated derivatives, and a massive parallelisation on a memory-distributed system towards solving extremely large-scale problems~\cite{darve2011fast,agullo2014task,agullo2016task,abduljabbar2017}.

\appendix

\section{Derivation of (\ref{eq:layer_disc}) from (\ref{eq:layer_disc_tmp2})}\label{s:simplify}

In the same way as \cite[Appendix~A]{takahashi2022}, the discretised vector potential $\ddot{\bm{A}}$ in (\ref{eq:layer_disc_A}) will be derived from that in (\ref{eq:layer_disc_tmp2_A}) via the concatenated current $\hat{\bm{J}}^\beta$ in (\ref{eq:hat_J}); the scalar and magnetic vector potentials in (\ref{eq:layer_disc_phi}) and (\ref{eq:layer_disc_P}), respectively, can be derived similarly. 

First, after introducing a shifted index $\beta':=\beta+\kappa$, the summation over $\beta'$ in (\ref{eq:layer_disc_A}) is split to three parts as follows:
\begin{eqnarray}
\ddot{\bm{A}}(\bm{x},t)
&=& \sum_{j=1}^{\Ne}\sum_{\kappa=0}^{d+1} \sum_{\beta'=\beta^*+\kappa}^{\alpha-1+\kappa} w^{\kappa,d} \ddot{\bm{A}}_j^{(\alpha-\beta')} \hat{J}_j^{\beta'-\kappa}\nonumber\\
&=& \sum_{j=1}^{\Ne}\sum_{\kappa=0}^{d+1} \left( \sum_{\beta'=\beta^*}^{\alpha-1}-\sum_{\beta'=\beta^*}^{\beta^*+\kappa-1}+\sum_{\beta'=\alpha}^{\alpha-1+\kappa}\right) w^{\kappa,d} \ddot{\bm{A}}_j^{(\alpha-\beta')} \hat{J}_j^{\beta'-\kappa},\label{eq:ddotA_tmp1}
\end{eqnarray}
where it is supposed that any summation $\sum_{i=s}^{e}$ is ignored if $s>e$. Then, the third summation $\sum_{\beta'=\alpha}^{\alpha-1+\kappa}$ disappears from the following reasons:
\begin{itemize}
\item For $\kappa=0$, the third summation becomes $\sum_{\beta'=\beta^*}^{\beta^*-1}$ and, thus, vanishes in accordance with the above convention. For this regard, the second summation also vanishes and the first one becomes exactly the original one.)

\item For $\kappa \ge 1$, the inequality $\alpha-\beta'\le 0$ is satisfied for any $\beta'\in[\alpha,\alpha-1+\kappa]$. Then, because $\ddot{\bm{A}}_j^{(\alpha-\beta')}$ is zero from the definition of $U$ in (\ref{eq:U}), the third summation vanishes.
\end{itemize}

Therefore, (\ref{eq:ddotA_tmp1}) can be written as follows:
\begin{eqnarray}
\ddot{\bm{A}}(\bm{x},t)
  &=& \sum_{j=1}^{\Ne} \left[ \sum_{\beta'=\beta^*}^{\alpha-1} \sum_{\kappa=0}^{d+1} w^{\kappa,d} \ddot{\bm{A}}_j^{(\alpha-\beta')} \hat{J}_j^{\beta'-\kappa}
- \underbrace{\sum_{\kappa=0}^{d+1} \sum_{\beta'=\beta^*}^{\beta^*+\kappa-1} w^{\kappa,d} \ddot{\bm{A}}_j^{(\alpha-\beta')} \hat{J}_j^{\beta'-\kappa}}_{\displaystyle =:F} \right]\nonumber\\
&=& \sum_{j=1}^{\Ne} \left[ \left(\underbrace{\sum_{\beta'=\beta^*}^{\beta^*+d}}_{\displaystyle =:G}+\underbrace{\sum_{\beta'=\beta^*+d+1}^{\alpha-1}}_{\displaystyle =:H}\right) \sum_{\kappa=0}^{d+1} w^{\kappa,d} \ddot{\bm{A}}_j^{(\alpha-\beta')} \hat{J}_j^{\beta'-\kappa} - F \right], \label{eq:ddotA_tmp2}
\end{eqnarray}
where the summation over $\beta'$ was split into $G$ and $H$. Since $\sum_{\kappa=0}^{d+1}\sum_{\beta'=\beta^*}^{\beta^*+\kappa-1}$ is identical to $\sum_{\beta'=\beta^*}^{\beta^*+d} \sum_{\kappa=\beta'-\beta^*+1}^{d+1}$ in $F$, the subtraction $G-F$ in (\ref{eq:ddotA_tmp2}) can be computed as follows:
\begin{eqnarray*}
  G-F
  &=&\sum_{\beta'=\beta^*}^{\beta^*+d}\sum_{\kappa=0}^{d+1} w^{\kappa,d} \ddot{\bm{A}}_j^{(\alpha-\beta')} \hat{J}_j^{\beta'-\kappa}
  -\sum_{\beta'=\beta^*}^{\beta^*+d}\sum_{\kappa=\beta'-\beta^*+1}^{d+1} w^{\kappa,d} \ddot{\bm{A}}_j^{(\alpha-\beta')} \hat{J}_j^{\beta'-\kappa}\nonumber\\
  &=&\sum_{\beta'=\beta^*}^{\beta^*+d} \ddot{\bm{A}}_j^{(\alpha-\beta')} \left(\sum_{\kappa=0}^{d+1}w^{\kappa,d} \hat{J}_j^{\beta'-\kappa}-\sum_{\kappa=\beta'-\beta^*+1}^{d+1}w^{\kappa,d}\hat{J}_j^{\beta'-\kappa}\right)\nonumber\\
  &=&\sum_{\beta'=\beta^*}^{\beta^*+d} \ddot{\bm{A}}_j^{(\alpha-\beta')} \sum_{\kappa=0}^{\beta'-\beta^*} w^{\kappa,d} \hat{J}_j^{\beta'-\kappa}.
\end{eqnarray*}
Meanwhile, $H$ can be rewritten as follows:
\begin{eqnarray*}
  H=\sum_{\beta'=\beta^*+d+1}^{\alpha-1}\sum_{\kappa=0}^{d+1} w^{\kappa,d} \ddot{\bm{A}}_j^{(\alpha-\beta')} \hat{J}_j^{\beta'-\kappa}
   =\sum_{\beta'=\beta^*+d+1}^{\alpha-1}\ddot{\bm{A}}_j^{(\alpha-\beta')} \sum_{\kappa=0}^{d+1} w^{\kappa,d} \hat{J}_j^{\beta'-\kappa}.
\end{eqnarray*}
The combination of $G-F$ and $H$ yields
\begin{eqnarray}
  \ddot{\bm{A}}(\bm{x},t)
  =\sum_{j=1}^{\Ne}\left[ G-F+H \right]
  =\sum_{j=1}^{\Ne}\left[ \sum_{\beta=\beta^*}^{\alpha-1}\ddot{\bm{A}}_j^{(\alpha-\beta)} \hat{J}_j^{\beta} \right],\label{eq:ddotA_tmp3}
\end{eqnarray}
where the following boundary variable $\hat{J}_j^\beta$ is defined:
\begin{eqnarray*}
  \hat{J}_j^{\beta}
  &:=&\begin{cases}
    \displaystyle\sum_{\kappa=0}^{\beta-\beta^*}w^{\kappa,d}\hat{J}_j^{\beta-\kappa} & \text{for $\beta^*\le\beta\le\beta^*+d$}\\
    \displaystyle\sum_{\kappa=0}^{d+1}w^{\kappa,d}\hat{J}_j^{\beta-\kappa} & \text{for $\beta^*+d+1\le\beta\le\alpha-1$}
  \end{cases}\nonumber\\
  &=&\sum_{\kappa=0}^{\min(d+1,\beta-\beta^*)}w^{\kappa,d}\hat{J}_j^{\beta-\kappa}\quad\text{for $\beta^*\le\beta\le\alpha-1$},
\end{eqnarray*}
which is exactly the same as (\ref{eq:hat_J}).

Finally, the index $\beta$ in (\ref{eq:ddotA_tmp3}) is increased by one to yield the simplified expression in (\ref{eq:layer_disc_A}).

\section*{Acknowledgements}

This study was supported by the JSPS KAKENHI Grant number JP21H03454.


\end{document}